\documentclass{cernyrep} 

\usepackage{texnames}
\usepackage[T1]{fontenc}
\begin{document}

\def\ang{\,{\rm\AA}}
\def\flux{\,{\rm erg\,cm^{-2}\,arcsec^{-2}\,\AA^{-1}\,s^{-1}}}
\def\GeV{\,{\rm GeV}}
\def\TeV{\,{\rm TeV}}
\def\k{\bf k}
\def\x{\bf x}
\def\gev{\,{\rm GeV}}
\def\H{\,{\cal  H}}
\def\keV{\,{\rm keV}}
\def\MeV{\,{\rm MeV}}
\def\sec{\,{\rm sec}}
\def\bi#1{\hbox{\boldmath{$#1$}}}
\newcommand{\Aa}{\ensuremath{\frac{a^{\prime}}{a}}}

\newcommand{\bea}{\begin{eqnarray}}
\newcommand{\eaa}{\end{eqnarray}}
\newcommand{\be}{\begin{equation}}
\newcommand{\ee}{\end{equation}}

\renewcommand{\vec}[1]{\bmath{#1}}
\newcommand{\ba}{\begin{eqnarray}}
\newcommand{\ea}{\end{eqnarray}}
\newcommand{\brr}{\begin{array}}
\newcommand{\err}{\end{array}}
\newcommand{\bc}{\begin{center}}
\newcommand{\ec}{\end{center}}
\renewcommand{\hm}{\,h^{-1}{\rm Mpc}}
\newcommand{\hk}{\,h^{-1}{\rm kpc}}
\newcommand{\msun}{\,h^{-1}M_\odot}
\newcommand{\hMpc}{\mbox{$h^{-1}{\rmn{Mpc}}~$}}
\newcommand{\rvir}{\mbox{$R_{\rmn{vir}}$}}
\newcommand{\mvir}{\mbox{$M_{\rmn{vir}}$}}
\newcommand{\tmw}{\mbox{$T_{\rmn{mw}}$}}
\newcommand{\hMpcI}{\mbox{$h\,{\rmn{Mpc}}^{-1}$}}
\newcommand{\lb}{{\left<\right.}}
\newcommand{\rb}{{\left.\right>}}
\newcommand{\lum}{\,{\rm erg\,s^{-1}}}
\newcommand{\vel}{\,{\rm km\,s^{-1}}}
\newcommand{\lt}{$L_X$--$T$$~$}
\newcommand{\bx}{\rm{\bf x}}
\newcommand{\by}{\rm{\bf y}}
\newcommand{\bk}{\rm{\bf k}}
\newcommand{\bv}{\rm{\bf v}}
\newcommand{\bu}{\rm{\bf u}}
\newcommand{\br}{\rm{\bf r}}
\newcommand{\tv}{t_{\rm vir}}
\newcommand{\lun}{\hbox{\ erg s$^{-1}$} }
\newcommand{\fun}{\hbox{\ erg cm$^{-2}$ s$^{-1}$} }
\def\arcmin{\hbox{$^\prime$}}
\def\arcsec{\hbox{$^{\prime\prime}$}}

\newcommand{\mincir}{\raise
  -2.truept\hbox{\rlap{\hbox{$\sim$}}\raise5.truept \hbox{$<$}\ }}
\newcommand{\magcir}{\raise
  -2.truept\hbox{\rlap{\hbox{$\sim$}}\raise5.truept \hbox{$>$}\ }}
\newcommand{\siml}{\raise
  -2.truept\hbox{\rlap{\hbox{$\sim$}}\raise5.truept \hbox{$<$}\ }}
\newcommand{\simg}{\raise
  -2.truept\hbox{\rlap{\hbox{$\sim$}}\raise5.truept \hbox{$>$}\ }}

\newcommand{\Ab}{\ensuremath{\Big(\frac{a^{\prime}}{a}\Big)^2}}
\newcommand{\Ac}{\ensuremath{\frac{a^{\prime\prime}}{a}}}
\newcommand{\Dp}{\ensuremath{\delta^{(1)}}}
\newcommand{\Ds}{\ensuremath{\delta^{(2)}}}
\newcommand{\La}{\ensuremath{\partial_i \,\partial^i}}
\newcommand{\deu}[1]{\ensuremath{{\delta #1}^{}}}
\newcommand{\ded}[1]{\ensuremath{{\delta #1}^{(2)}}}
\newcommand{\ze}[1]{\ensuremath{#1}^{(0)}}
\def\Gyr{\,{\rm Gyr}}
\def\yr{\,{\rm yr}}
\def\rcm{\,{\rm cm}}
\def\pc{\,{\rm pc}}
\def\kpc{\,{\rm kpc}}
\def\Mpc{\,{\rm Mpc}}
\def\mpc{\,{\rm Mpc}}
\def\eV{{\,\rm eV}}
\def\ev{{\,\rm eV}}
\def\erg{{\,\rm erg}}
\def\cmm2{{\,\rm cm^{-2}}}
\def\cm2{{\,{\rm cm}^2}}
\def\cmm3{{\,{\rm cm}^{-3}}}
\def\gcmm3{{\,{\rm g\,cm^{-3}}}}
\def\kms{\,{\rm km\,s^{-1}}}
\def\HO{{100h\,{\rm km\,sec^{-1}\,Mpc^{-1}}}}
\def\mpl{{m_{\rm Pl}}}
\def\pl{{\rm Pl}}
\def\mpp{{m_{\rm Pl,0}}}
\def\trh{T_{\rm RH}}
\def\g{\tilde g}
\def\R{{\cal R}}
\def\km{\rm \,km}
\def\yrs{\rm \,yrs}
\def\trh{T_{\rm RH}}

%def\baselinestretch{1.4}
\def\VEV#1{\left\langle #1\right\rangle}
\def\la{\mathrel{\mathpalette\fun <}}
\def\ga{\mathrel{\mathpalette\fun >}}
\def\fun#1#2{\lower3.6pt\vbox{\baselineskip0pt\lineskip.9pt
  \ialign{$\mathsurround=0pt#1\hfil##\hfil$\crcr#2\crcr\sim\crcr}}}

\title{Particle cosmology}
 
\author{A. Riotto}

\institute{CERN, Geneva, Switzerland}

\maketitle % this produces the title block

\begin{abstract}
In these lectures the present status of the so-called standard cosmological model, based on the hot 
Big Bang theory and the inflationary paradigm is reviewed. Special emphasis is given to the origin of the cosmological
perturbations we see today under the form of the cosmic microwave background anisotropies and the large scale structure and
to the dark matter.

\end{abstract}
 
\section{Introduction}

There are fundamental questions we are on the edge of answering: what is the origin of our universe? Why is the universe 
so homogeneous and isotropic on large scales? What are the origins of dark matter and dark energy? What  is the fate
of our universe?  While  these lectures will certainly not be able to give  definite answers to them, we shall
try to provide the students with some tools they might find useful in order to solve  these 
overwhelming mysteries themselves.

These lectures will contain a short review of the standard Big Bang model; a rather long
discussion of the inflation paradigm with particular emphasis on the possibility that the cosmological seeds
originated from a period of primordial acceleration; the physics of the Cosmic Microwave Background (CMB) 
anisotropies,  and a short discussion of the dark matter

 Since these lectures were delivered at  a school, 
we shall not provide an exhaustive list of
references to original material, but refer to several basic cosmology books and reviews where students can find the
references to the original material \cite{abook,kolbbook,kolb,tur,linde,Liddle,llbook,lr,riottoictp,cmb,copeland}.

\section {Basics of the Big Bang model}
The key starting point of the description of our universe is the fact that the latter looks homogeneous and isotropic on large scales. For such a reason the  standard cosmology is based upon the maximally spatially symmetric
Friedmann--Robertson--Walker (FRW) line element
\begin{equation}
ds^2 = -dt^2 +a(t)^2\left[ {dr^2\over 1-kr^2} +r^2
        (d\theta^2 + \sin^2\theta\,d\phi^2 ) \right]\, ;
\label{metric}
\end{equation}
Here $a(t)$ is the cosmic-scale factor; the curvature is parametrized by  $R_{\rm curv}\equiv
a(t)|k|^{-1/2}$ and the parameter $k$ can be chosen to acquire the values $= -1,
0, 1$  The coordinates are 
 {\it co-moving} coordinates and It is important to point out that physical separations  between freely moving particles
scale as $a(t)$.  The momenta of freely propagating particles
scale like  as $a(t)^{-1}$,while 
wavelength of  photons stretches as $a(t)$.  Correspondingly,  the redshift
suffered by a photon emitted from a distant galaxy
$1+z = a_0/a(t)$.

The evolution of the scale factor $a(t)$ is governed by Einstein equations
\begin{equation}
R_{\mu\nu}-\frac{1}{2}\,R\,g_{\mu\nu}\equiv G_{\mu\nu}=8\pi G\, , 
T_{\mu\nu}
\end{equation}
where $R_{\mu\nu}$ $(\mu,\nu=0,\cdots 3)$ 
is the Riemann tensor and $R$ is the Ricci scalar
constructed via the metric (\ref{metric}) \cite{kolbbook,kolb},  and  
$T_{\mu\nu}$ is the energy-momentum tensor. $G=m_{\rm Pl}^{-2}$ is the Newton constant.
Under the hypothesis
of homogeneity and isotropy, we can always write the energy-momentum
tensor under the form $T_{\mu\nu}={\rm diag}\left(\rho,P,P,P\right)$
where $\rho$ is the energy density of the system and $P$ its pressure.
They are functions of time.

The evolution of the cosmic-scale factor is governed
by the Friedmann equation
\begin{equation}
H^2 \equiv \left({\dot a \over a}\right)^2 =
        {8\pi G \rho \over 3} - {k\over a^2}\,, 
\label{fr1}
\end{equation}
where $\rho$ is the total energy density of the
universe, matter, radiation, vacuum energy, and so on.

Differentiating wrt to time both members of Eq. (\ref{fr1}) and using the 
the mass conservation equation
\begin{equation}
\label{mass}
\dot{\rho} +3H(\rho +P) =0\, ,
\end{equation}
we find the equation for the acceleration of the scale factor
\begin{equation}
\frac{\ddot{a}}{a} = - \frac{4\pi G}{3}
(\rho +3P).
\label{fr2}
\end{equation}
Combining Eqs. (\ref{fr1}) and (\ref{fr2}) we find
\begin{equation}
\dot H=-4\pi G\left(\rho+P\right).
\label{ll}
\end{equation}
The evolution of the energy
density of the universe is governed by
\begin{equation}
d(\rho a^3) = -P d\left(a^3\right);
\end{equation}
which is the first law of thermodynamics for
a fluid in the expanding universe.
(In the case that the stress energy of the universe is comprised
of several, non-interacting components, this relation applies
to each separately; {\it e.g.}, to the matter and radiation separately
today.)  For $P=\rho /3$, ultra-relativistic matter,
$\rho \propto a^{-4}$ and $a\sim t^{\frac{1}{2}}$; 
for $P=0$, very nonrelativistic
matter, $\rho \propto a^{-3}$ and $a\sim t^{\frac{2}{3}}$;
 and for $P=-\rho$, vacuum
energy, $\rho = \,$const.  If the rhs of the Friedmann
equation is dominated by a fluid
with equation of state $P = w \rho$, it follows
that $\rho \propto a^{-3(1+w )}$
and $a\propto t^{2/3(1+w )}$.

 Friedmann equations  relate the
curvature of the universe to the energy density and
expansion rate:
\begin{equation}
 \Omega -1={k \over a^2H^2}\, ; \qquad
\Omega = {\rho\over \rho_{\rm crit}}\, ;
\label{curvature}
\end{equation}
and the critical density today $\rho_{\rm crit}
= 3H^2 /8\pi G = 1.88h^2\gcmm3 \simeq 1.05\times 10^{4}
\eV \cmm3$.  The curvature radius of the universe is related
to the Hubble radius and $\Omega$ by
\begin{equation}
R_{\rm curv} = {H^{-1}\over |\Omega -1|^{1/2}}\, .
\label{curv}
\end{equation}
In physical terms, the curvature radius sets the scale for
the size of spatial separations where
the effects of curved space become
pronounced.  

The energy content of the universe consists of matter
and radiation (today, photons and neutrinos).  Since
the photon temperature is accurately known,
$T_0=2.73\pm 0.01\,$K, the
fraction of critical density contributed by radiation
is also accurately known:  $\Omega_{R}h^2 = 4.2 \times
10^{-5}$, where $h=0.72\pm 0.07$ is the present Hubble rate in units of 
$100$ km 
${\rm s}^{-1}$ ${\Mpc}^{-1}$ \cite{wmap5}.  The remaining content of the 
universe is 
another 
matter. Rapid progress has been made recently toward the measurement of 
cosmological parameters \cite{triangle}. We know by now that  the universe  
is spatially flat; 
accelerating; comprised of one third dark matter
and two thirds a new 
form of dark energy \cite{wmap5} 

$$\Omega_{0} 
=1.00^{+0.07}_{-0.03}\,,
$$
meaning that the present universe is spatially flat (or at least very close
to being flat). Restricting to    $\Omega_0=1$, the dark matter density
is given by \cite{wmap5}

$$ 
\Omega_{\rm DM}h^2 
=0.11^{+0.0034}_{-0.059}\,,
$$
and a baryon density

$$
\Omega_B  = 0.045\pm 0.0015,
$$
while the Big Bang nucleosynthesis estimate is $\Omega_B  
h^2=0.019\pm
     0.002.$ Substantial dark (unclustered) energy is inferred:

$$
\Omega_{\rm DE}  
\approx 0.72 \pm 0.015\, .
$$ 
What is most
relevant for us is that  this universe was apparently born from a burst of rapid 
expansion, inflation, during which quantum noise was stretched to 
astrophysical size 
seeding cosmic structure. This is exactly the phenomenon we want to address 
in part of these lectures.

\subsection{The early, radiation-dominated universe}

In an expanding universe  the energy density
in matter decreases as $a^{-3}$, and that
in radiation as $a^{-4}$ at early times the
universe was radiation dominated.

Denoting the epoch of matter and radiation equality
by subscript `EQ,' and using $T_0=2.73\,$K, it follows that
\begin{equation}
T_{\rm EQ} = 5.62 (\Omega_0 h^2)\eV\, .
\end{equation}
At early times, when the universe is radiation dominated, the expansion rate 
determined by the temperature of the universe and
the number of relativistic degrees of freedom
\begin{equation}
\rho_{\rm rad} = g_*(T){\pi^2T^4 \over 30}; \qquad
H\simeq 1.67g_*^{1/2} T^2 /\mpl ;
\end{equation}
\begin{equation}
\Rightarrow a\propto t^{1/2}; \qquad
t \simeq 2.42\times 10^{-6} g_*^{-1/2}(T/\GeV )^{-2}\,{\rm s} \,;
\end{equation}
where $g_* (T)$ counts the number of ultra-relativistic
degrees of freedom ($\approx$ the sum of the internal
degrees of freedom of particle species much less massive
than the temperature) and $\mpl \equiv G^{-1/2} = 1.22
\times 10^{19}\GeV$ is the Planck mass.

A quantity of importance related to $g_*$ is the
entropy density in relativistic particles,
$$s= {\rho +p \over T} = {2\pi^2\over 45}g_* T^3 ,$$
and the entropy per co-moving volume,
$$S \ \  \propto\ \  a^3 s\ \  \propto\ \   g_*a^3T^3 .$$

Since in thermal equilibrium the entropy per co-moving
volume $S$ remains constant, we get that the temperature and scale
factor are related by
\begin{equation}
T\propto g_*^{-1/3}a^{-1},
\end{equation}
which for $g_*=\,$const leads to the familiar $T\propto a^{-1}$.
Also, in our present Hubble volume in
a very physical way:  by the entropy it contains,
\begin{equation}
S_{U} = {4\pi\over 3}H_0^{-3}s \simeq 10^{90},
\end{equation}
a huge number indeed, which will play a crucial role in the following.

\subsection{The concept of particle horizon}

In a FRW cosmology 
photons travel on geodesics satisfying the equation
 $dr=dt/a(t)$. This means that in a time $t$ photons travel a distance
 
 \begin{eqnarray}
R_H(t) & = & a(t)\int_0^t {dt^\prime\over a(t^\prime )} \nonumber\\
       & = & \frac{t}{(1-n)} = n\,\frac{H^{-1}}{(1-n)}\sim
H^{-1} \qquad {\rm for}\ a(t)
        \propto t^n, \ \ n<1 .
\end{eqnarray}
Using the conformal time $d\tau=dt/a$, the particle horizon becomes
\be
R_H(t)=a(\tau)\int_{\tau_0}^\tau\, d\tau,
\end{equation}
where $\tau_0$ indicates the conformal time corresponding to $t=0$.
This quantity is very close to the 
 Hubble radius during radiation or matter periods.
\footnote{A word of caution:  in inflationary models
the horizon and Hubble radius are exponetially different .}.

A physical length scale $\lambda$ is within the horizon when the following condition is satisfied: 
if $\lambda<R_{H}\sim H^{-1}$. Setting the  length scale to be 
$\lambda=\lambda=2\pi a/k$, 
we shall have the following
rule
\begin{center}
\begin{tabular}{|p{13 cm}|}
\hline
%\\
\begin{eqnarray}
\frac{k}{aH}&\ll& 1 \Longrightarrow {\rm SCALE}~~\lambda~~
{\rm OUTSIDE}~~{\rm THE}~~{\rm HORIZON}\nonumber\\
\frac{k}{aH}&\gg& 1 \Longrightarrow {\rm SCALE}~~\lambda~~
{\rm WITHIN}~~{\rm THE}~~{\rm HORIZON}\nonumber
\end{eqnarray}
%\\
\\
\hline
\end{tabular}
\end{center}

\section{The shortcomings of the standard Big Bang theory}

The most accurate measurement of the temperature
and spectrum is that by the WMAP5 instrument on the
COBE satellite which determined its temperature to be
$2.726\pm 0.01\,$K \cite{wmap5}.  
The length corresponding to our present Hubble radius (which is
approximately the radius of our observable universe) at the time
of last scatteringwas
$$
\lambda_H(t_{\rm LS})=R_H(t_0) \left(\frac{a_{\rm LS}}{a_0}
\right)=R_H(t_0) \left(\frac{T_{0}}{T_{\rm LS}}\right).
$$
On the other hand, during the matter-dominated period, 
the Hubble length  decreased with a different law
$$
H^2\propto \rho_{M} \propto a^{-3} \propto T^{3}.
$$
At last-scattering
$$
H_{LS}^{-1}=R_H(t_0)\left( \frac{T_{LS}}{T_0} \right)^{-3/2}\ll R_H(t_0).
$$
The length corresponding to our present Hubble radius was much
larger that the horizon at that time. This can be by shown comparing
the volumes corresponding to these two scales

\begin{equation}
\frac{\lambda^3_{H}(T_{LS})}{H_{LS}^{-3}}=
\left(\frac{T_0}{T_{LS}}\right)^{-\frac{3}{2}}\approx 10^6.
\label{pp}
\end{equation}
%%%%%%%%%%%%%%%%%%%%%%%%%%%%%%%%%%%%%%%%%%%%%%%%%%%%%%%%%%%%%%%%
\begin{figure}
\centering\includegraphics[width=.5\linewidth]{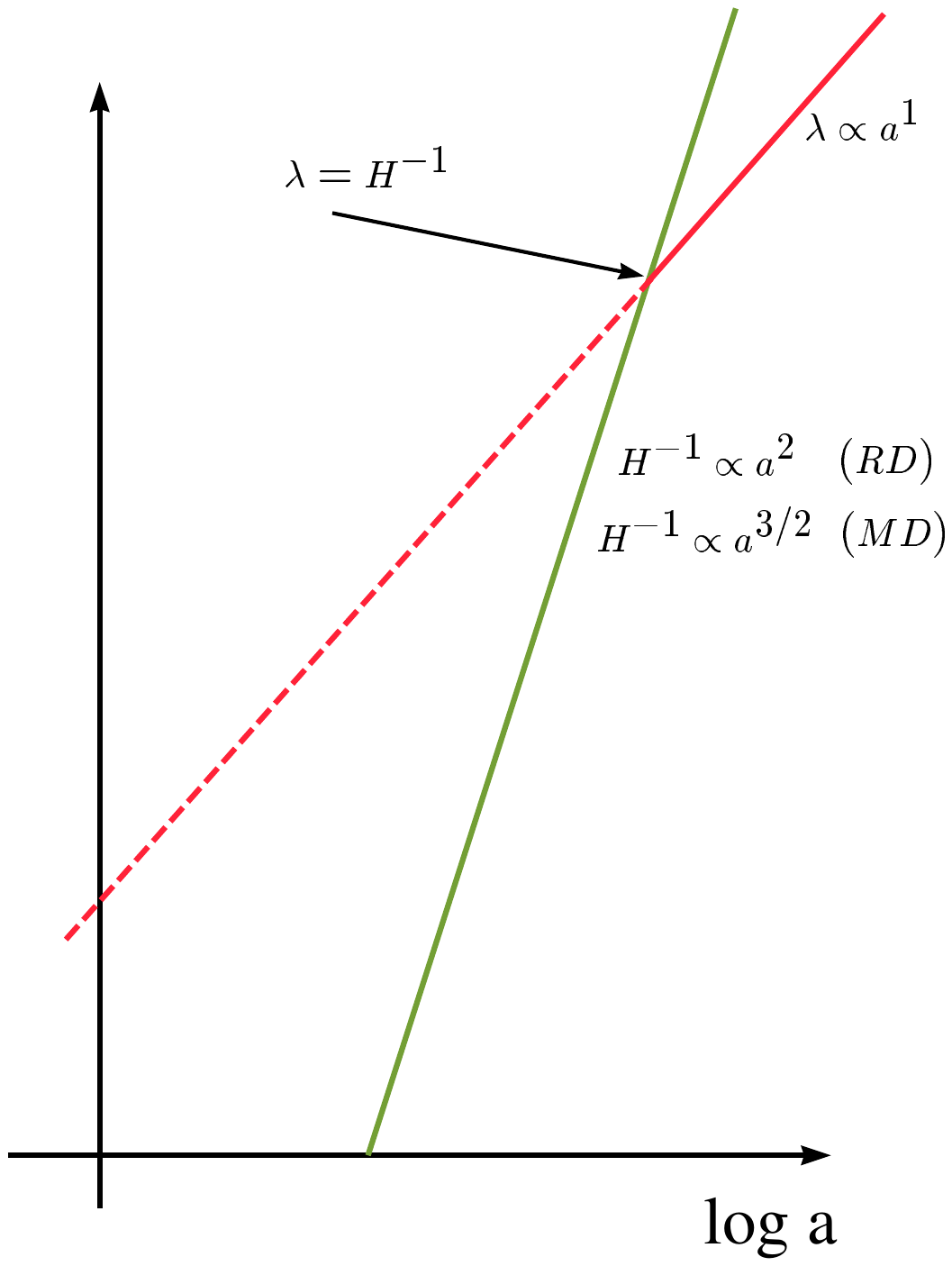}
\caption{The horizon scale (solid line) and a physical scale $\lambda$
(dashed line) as function of the scale factor $a$}
\label{normal}
\end{figure}
%%%%%%%%%%%%%%%%%%%%%%%%%%%%%%%%%%%%%%%%%%%%%%%%%%%%%%%%%%%%%%%%
There were $\sim 10^6$ casually
disconnected regions within the volume that now corresponds to our
horizon! 
It is difficult to
come up with a process other
than an early hot and dense phase in the history
of the universe that would lead to a precise
black body  for a bath of photons which
were causally disconnected the last time  they interacted with the
surrounding plasma.

The horizon problem is well represented by Fig. \ref{normal}
where the solid line indicates the horizon scale and the dashed line any
generic physical length scale $\lambda$. Suppose, indeed, that $\lambda$
indicates the distance between two photons we detect today. From
Eq. (\ref{pp}) we discover that at the time of emission (last-scattering)
the two photons could not talk to each other, the dashed line is above the
solid line  \cite{kolbbook}.
%\begin{figure}[htb]
%\centerline{\hbox{ \psfig{figure=normal.ps} }}
%\caption{\footnotesize Explanation \label{normal}}
%\end{figure}
 \cite{kolbbook}.
There is another main point to mention with the horiton problem and it is related to the inhomogeneities. We know that 
the temperature anisotropy on the angular scale subtended
by that length scale,
\begin{equation}
\left({\delta T \over T}\right)_\theta 
\approx
\left( {\delta\rho\over \rho}\right)_{\lambda},
\end{equation}
where the scale $\lambda \sim 100h^{-1}\Mpc(\theta /{\rm deg})$
subtends an angle $\theta$ on the last-scattering
surface.  This is known as the Sachs--Wolfe effect \cite{SW,scott}.
We shall come back to this piece of physics.
%%%%%%%%%%%%%%%%%%%%%%%%%%%%%%%%%%%%%%%%%%%%%%%%%%%%%%%%%%%%%%%%
\begin{figure}
\centering\includegraphics[width=.5\linewidth]{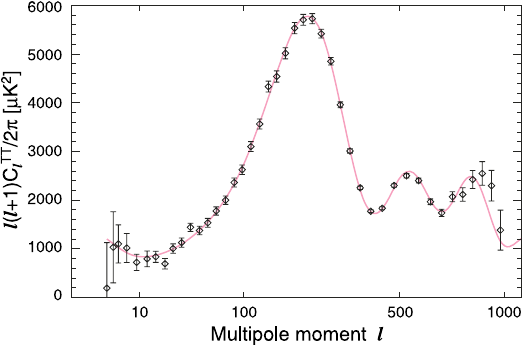}
\caption{The CMBR anisotropy as function of $\ell$ (from Ref. \cite{wmap5})}
\label{f1}
\end{figure}
%%%%%%%%%%%%%%%%%%%%%%%%%%%%%%%%%%%%%%%%%%%%%%%%%%%%%%%%%%%%%%%%
The
temperature anisotropy is commonly expanded  in spherical harmonics
\begin{equation}
\frac{\Delta T}{T}(x_0,\tau_0,{\bf n})=\sum_{\ell m}
a_{\ell,m}(x_0)Y_{\ell m}({\bf n}),
\end{equation}
where $x_0$ and $\tau_0$ are our position and the preset time, respectively,
 ${\bf n}$ is the
direction of observation, $\ell'$s are the
different multipoles  and\footnote{An alternative definition is $C_\ell=
\langle \left|a_{\ell m}\right|^2\rangle=\frac{1}{2\ell +1}
\sum_{m=-\ell}^{\ell}
\left|a_{\ell m}\right|^2$.}
\begin{equation}
\langle a_{\ell m}a^*_{\ell'm'}\rangle=\delta_{\ell,\ell'}\delta_{m,m'} C_\ell,
\end{equation}
where the deltas are due to the fact that the process that created
the anisotropy is statistically isotropic. 
%\begin{figure}[htb]
%\centerline{\hbox{ \psfig{figure=cobe.eps} }}
%\caption{\footnotesize A plot for illustration.  \label{figcobe}}
%\end{figure}
The $C_\ell$'s are the so-called CMB power spectrum.
For homogeneity and isotropy, the $C_\ell$'s are neither a function
of $x_0$, nor of $m$.
The two-point correlation function is related to the $C_l$'s in
the following way
\begin{eqnarray}
\Big<\frac{\delta T({\bf n})}{T}\frac{\delta T({\bf n}')}{T}\Big>&=&
\sum_{\ell\ell'mm'}\langle a_{\ell m}a^*_{\ell'm'}\rangle
Y_{\ell m}({\bf n})Y^*_{\ell'm'}({\bf n}')\nonumber\\
&=&\sum_\ell C_\ell \sum_m
Y_{\ell m}({\bf n})Y^*_{\ell m}({\bf n}')=\frac{1}{4\pi}\sum_\ell (2\ell+1) 
C_\ell
P_\ell(\mu={\bf n}\cdot{\bf n}')
\label{j}
\end{eqnarray}
where we have used the addition theorem for the spherical
harmonics, and $P_\ell$ is the Legendre polynom of order $\ell$. 
In expression (\ref{j}) the expectation value is an ensemble average. It 
can be regarded as an average over the possible observer positions, but
not in general as an average over the single sky we observe, because of the
cosmic variance\footnote{The usual hypothesis is that we observe a typical
realization of the ensemble. This means that we expect  the difference 
between the observed values $|a_{\ell m}|^2$ and the 
ensemble averages $C_\ell$ to be of the order of the mean-square deviation
of  $|a_{\ell m}|^2$ from $C_\ell$. The latter is called
cosmic variance and, because we are dealing with a Gaussian distribution, it is
equal to $2C_\ell$ for each multipole $\ell$. For a single $\ell$, averaging 
over the $(2\ell +1)$ values of $m$ reduces the cosmic variance
by a factor $(2\ell +1)$, but it remains a serious limitation for low 
multipoles.}. WMAP5 data are given in Fig. \ref{f1}.

Let us now consider the last scatteringsurface. In co-moving coordinates
the latter is `far' from us a distance equal to
\be
\int_{t_{\rm LS}}^{t_0}\,\frac{dt}{a}=\int_{\tau_{\rm LS}}^{\tau_0}\,
d\tau=\left(\tau_0-\tau_{\rm LS}\right).
\end{equation}
A given co-moving 
scale $\lambda$  is therefore projected on the last scatteringsurface
sky on an angular scale
\begin{equation}
\theta \simeq \frac{\lambda}{\left(\tau_0-\tau_{\rm LS}\right)},
\end{equation}
where we have neglected tiny curvature effects.
Consider now
that the scale $\lambda$ is of the order of the co-moving sound horizon at the 
time of last-scattering, $\lambda\sim c_s\tau_{\rm LS}$, where
$c_s\simeq 1/\sqrt{3}$ is the sound velocity at which photons
propagate in the plasma at the last-scattering.
This corresponds
to an angle 
\be
\theta\simeq c_s\frac{\tau_{\rm LS}}{\left(\tau_0-\tau_{\rm LS}\right)}\simeq
c_s\frac{\tau_{\rm LS}}{\tau_0},
\end{equation}
where the last passage has been performed knowing that $\tau_0\gg
\tau_{\rm LS}$. Since the universe is matter-dominated from the
time of last scatteringonwards, the scale factor has the following
behaviour:  $a\sim T^{-1}\sim 
t^{2/3}\sim \tau^2$.
The angle $\theta_{\rm HOR}$ 
subtended by the sound horizon on the last-scattering
surface then becomes
\be
\theta_{\rm HOR}
\simeq c_s\left(\frac{T_0}{T_{\rm LS}}\right)^{1/2}\sim 1^\circ,
\end{equation}
where we have used $T_{\rm LS}\simeq 0.3$ eV and $T_0\sim 10^{-13}$ GeV.
This corresponds to a multipole $\ell_{\rm HOR}$
 
\begin{equation}
\ell_{\rm HOR}=\frac{\pi}{\theta_{\rm HOR}}\simeq 200\, .
\end{equation}
From these estimates we conclude that  
two photons which on the last scatteringsurface were separated
by an angle larger than $\theta_{\rm HOR}$, corresponding to
multipoles smaller than $\ell_{\rm HOR}\sim 200$, were not in causal
contact. 
On the other hand, 
from Fig. \ref{f1} it is clear that small anisotropies, of the 
{\it same} order of magnitude $\delta T/T\sim 10^{-5}$ are present at $\ell\ll
200$. We conclude that one of the striking features of the CMB 
fluctuations is that they appear to be non-causal.

As can be seen in Fig. \ref{normal}, in the
standard cosmology the physical
size of a perturbation, which grows as the scale factor,
begins larger than the horizon and, relatively late
in the history of the universe, crosses inside the horizon.
This precludes a causal microphysical explanation for
the origin of the required density perturbations.

From the considerations made so far, it appears that solving the
horizon problem of the standard Big Bang theory requires 
that the universe go through a primordial period during which
the physical scales $\lambda$ evolve faster than the horizon scale $H^{-1}$.

\begin{figure}
\centering\includegraphics[width=.5\linewidth]{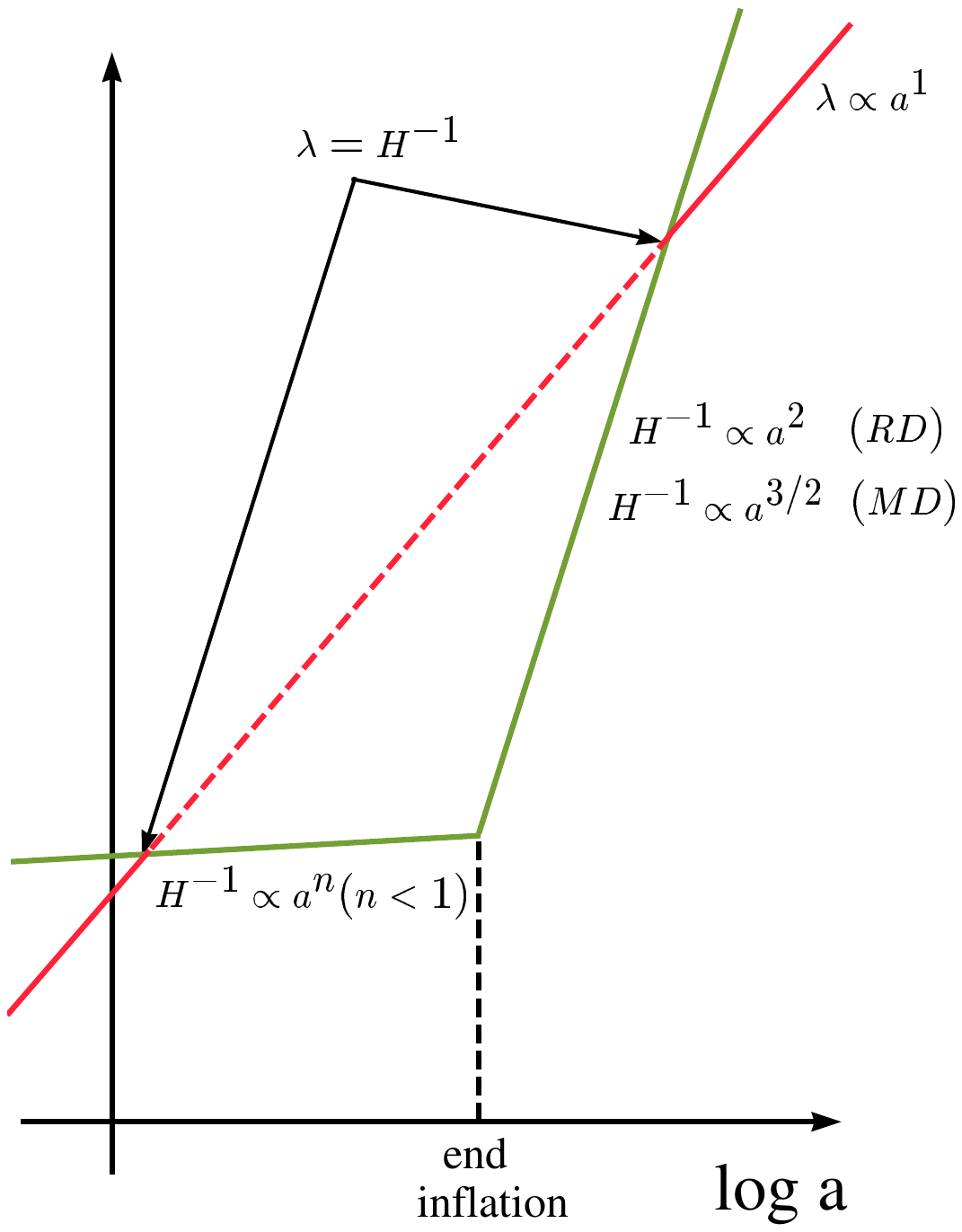}
\caption{The behaviour of a generic scale $\lambda$ and the horizon scale
$H^{-1}$ in the standard inflationary model}
\end{figure}

If there is period during which physical length scales grow faster
than $H^{-1}$, length scales $\lambda$ 
which are within the horizon today, $\lambda<H^{-1}$ (such
as the distance between two detected photons) and 
were outside the
horizon for some period,  $\lambda>H^{-1}$ (for 
instance at the time of last scatteringwhen the
two photons were emitted), had a chance to be within the horizon
at some primordial epoch,  $\lambda<H^{-1}$ again, see Fig. 3. If this happens,
the homogeneity and the isotropy of the CMB can easily be explained:
photons that we receive today and were emitted from the last scattering
surface from causally disconnected regions have the same temperature
because they had a chance to `talk' to each other at some primordial
stage of the evolution of the universe.

The second condition can easily be expressed as a condition on the
 scale factor $a$. Since a given scale $\lambda$ scales like
$\lambda\sim a$ and $H^{-1}=a/\dot a$, we need to impose
that there is a period during which
$$
\left(\frac{\lambda}{H^{-1}}\right)^{\cdot}=\ddot a>0\, .
\label{fund}
$$ 
We can therefore introduce
the following rigorous definition: an  inflationary stage  is 
a period of the universe during which the latter accelerates
\begin{center}
\begin{tabular}{|p{13.0 cm}|}
\hline
%\\
$$
{\rm INFLATION}~~~\Longleftrightarrow~~~\ddot a>0.
$$
%\\
\\
\hline
\end{tabular}
\end{center}
\vskip 0.2cm

{\it \underline{Comment}:}~Let us stress that during such an accelerating  phase 
the universe expands {\it adiabatically}. This means that during inflation
one can exploit the usual FRW equations (\ref{fr1}) and (\ref{fr2}). 
It must be 
clear therefore that the non-adiabaticity condition is satisfied not during 
inflation, but during the phase transition between the end of inflation
and the beginning of the radiation-dominated phase. At this transition phase
a large entropy is generated under the form of relativistic degrees of freedom:
the Big Bang has taken place.

\section{The standard inflationary universe}
Some parts which follow are taken from \cite{kolb} and \cite{linde}. We thank E.W. Kolb fand A. Linde or granting permission.
From the previous section we have learned that 
an accelerating stage during the primordial phases  of the evolution of
the universe
might be able to solve the horizon problem. From Eq. (\ref{fr2})
we learn that 
$$
\ddot a>0 \Longleftrightarrow (\rho +3P)<0\,.
\label{gg}
$$
An accelerating period is obtainable only if the overall pressure $p$ 
of the universe is negative: $p<-\rho/3$. Neither 
a radiation-dominated
phase nor a matter-dominated phase (for which $p=\rho/3$ and $p=0$, 
respectively) satisfy such a condition. Let us postpone for the time being
the problem of finding a `candidate' able to provide the condition
$P<-\rho/3$. For sure, 
inflation  is a phase of the history of the universe
occurring before the era of nucleosynthesis ($t \approx 1$ s, $T
\approx 1$ MeV) during which the light elements abundances were
formed. This is because nucleosynthesis 
is the earliest epoch  from which we have
experimental data  and they are in agreement
with the predictions of the standard Big Bang theory. However,  
the thermal history of the universe before 
the epoch of nucleosynthesis is unknown. 

In order to study the properties of the period of inflation, we assume the
extreme condition $p=-\rho$ which considerably simplifies the
analysis. A period of the universe during which
$P=-\rho$ is called the de~Sitter stage. 
By inspecting Eqs. (\ref{fr1}) and 
(\ref{mass}),
we learn that during the de Sitter phase 
\begin{eqnarray}
\rho&=&~~{\rm constant}\, ,\nonumber\\
H_I&=&~~{\rm constant}\, ,\nonumber
\end{eqnarray}
where we have indicated by $H_I$ the value of the Hubble rate during inflation.
Correspondingly, solving Eq. (\ref{fr1}) gives
\begin{equation}
a=a_i\, e^{H_I(t-t_i)},
\end{equation}
where $t_i$ denotes the time at which inflation starts.
Let us now see how such a period of exponential expansion
takes care of the shortcomings of the standard Big Bang Theory\footnote{
Despite the fact that the growth of the scale factor is exponential
and the expansion is {\it superluminal}, this is not
in contradiction with what is dictated by relativity. Indeed, it is the
spacetime itself which is progating so fast and not a light signal in it.}.

\subsection{Inflation and the horizon problem}

During the  inflationary (de Sitter) epoch the horizon scale $H^{-1}$  is
constant. If inflation lasts long enough, all the physical scales
that have left  the horizon during the radiation-dominated or 
matter-dominated phase can
re-enter the horizon in the past: this is 
because such scales  are exponentially reduced.
As we have seen in the previous section, 
this explains both the problem of the homogeneity of CMB
and the initial condition problem of small
cosmological perturbations.
Once the physical length is within the horizon,
microphysics can act, the universe can be made
approximately homogeneous and the primeval inhomogeneities can
be created. 

Let us see how long inflation must
be sustained in order to solve the horizon problem.
Let $t_i$ and $t_f$ be, respectively, the time of beginning and
end of inflation. We can define the corresponding number of e-foldings $N$
\begin{equation}
N={\rm ln}\left[H_I(t_e-t_i)\right].
\end{equation}
A necessary condition to solve the horizon problem is that the
largest scale we observe today, the present horizon $H_0^{-1}$, was
reduced  during inflation to a value $\lambda_{H_0}(t_{i})$ 
smaller than the value of
horizon length $H_I^{-1}$ during inflation.
This gives
$$
\lambda_{H_0}(t_{i})=H^{-1}_0 \left(\frac{a_{t_f}}{a_{t_0}}
\right) \left(\frac{a_{t_i}}{a_{t_f}}\right)=
H_0^{-1} \left(\frac{T_0}{T_f}
\right) e^{-N}\la H_I^{-1},
$$
where we have neglected for simplicity the short period of matter-domination
and we have called $T_f$ the temperature at the end of inflation (to be 
indentified with  the reheating temperature $T_{RH}$ at the beginning of 
the radiation-dominated phase after inflation, see later).  
We get 
$$
N \ga \ln\left( \frac{T_0}{H_0} \right) - \ln\left( \frac{T_f}{H_I}
\right) \approx 67 + \ln\left( \frac{T_f}{H_I} \right).
$$
Apart from the logarithmic dependence, we obtain  $N \ga 70$.

\subsection{A prediction of inflation }

Since during inflation the
Hubble rate is constant
$$
\Omega -1 = \frac{k}{a^2H^2}\propto \frac{1}{a^2}\ .
$$
On the other hand it is easy to show that  
to reproduce a value of $(\Omega_0-1)$ of order of unity today,
the initial value of $(\Omega-1)$ at the beginning of the
radiation-dominated phase must be $\left|\Omega-1\right|\sim 10^{-60}$.
Since we identify the beginning of the radiation-dominated phase
with the beginning of inflation, we require
$$
\left|\Omega -1\right|_{t=t_{f}}\sim 10^{-60}.
$$
During inflation
\begin{equation}
\frac{\left|\Omega -1\right|_{t=t_{f}}}{\left|
\Omega -1\right|_{t=t_{i}}}= \left(\frac{a_i}{a_f}
   \right)^2 = e^{-2N}.
\label{fold}
  \end{equation}
Taking $\left|
\Omega -1\right|_{t=t_{i}}$ of order unity, 
it is enough to
require that $N \approx 70$. However, IF  the period of inflation lasts longer than 70 e-foldings
the present-day value of $\Omega_0$ will be equal to unity with  great
precision. One can say that a generic prediction of inflation is that
\begin{center}
\begin{tabular}{|p{13.0 cm}|}
\hline
%\\
$$
{\rm INFLATION}~~~\Longrightarrow~~~\Omega_0=1.
$$
%\\
\\
\hline\end{tabular}
\end{center}
This statement, however, must be taken {\it cum grano salis} and
properly specified. Inflation does not change the global geometric properties
of the space-time. If the universe is open or closed, it will
always remain flat or closed, independently from inflation. 
What inflation does is to magnify the radius of curvature 
$R_{\rm curv}$ defined in Eq. (\ref{curv}) 
so that locally
the universe is flat with a great precision. 
As we shall see, the current data on the
CMB anisotropies confirm this prediction.

\subsection{Inflation and the inflaton}

In the previous subsections we have described the various
advantages of having a period of accelerating phase. The latter
required $P<-\rho/3$. Now, we would like to show that this condition
can be attained by means of  a simple 
scalar field. We shall call  this field the {\it  inflaton} $\phi$.

The action of the inflaton field reads
\begin{equation}
S=\int d^4x\, \sqrt{-g}\,\mathcal{L}=\int\, d^4x\, \sqrt{-g}\,
\left[\frac{1}{2}
\partial_{\mu}\phi
\partial^{\mu}\phi +V(\phi)\right],
\end{equation}
where $\sqrt{-g}=a^3$ for the FRW metric (\ref{metric}).
From the Euler--Lagrange equations
\begin{equation}
\partial^{\mu}\frac{\delta(\sqrt{-g}\mathcal{L})}{\delta\,
\partial^{\mu}\phi}- \frac{\delta(\sqrt{-g}\mathcal{L})}{\delta
\phi}=0\,, 
\end{equation}
we obtain
\begin{equation}
\ddot{\phi}+ 3H\dot{\phi}-\frac{\nabla^2\phi}{a^2}+V'(\phi)=0\, ,
\label{nabla}
\end{equation}
where $V'(\phi)=\left(dV(\phi)/d\phi\right)$. Note, in particular, the
appearance of the friction term $3H\dot{\phi}$: a scalar field
rolling down its potential suffers a friction due to the
expansion of the universe.

We can write the energy momentum tensor of the scalar field
$$
T_{\mu\nu}=\partial_{\mu}\phi \partial_{\nu}\phi
-g_{\mu\nu}\, \mathcal{L}\, .
$$
The corresponding energy density $\rho_\phi$ and pressure density $P_\phi$ 
are
\begin{eqnarray}
T_{00}=\rho_{\phi}=\frac{\dot{\phi}^2}{2} + V(\phi)+ 
\frac{(\nabla \phi)^2}{2a^2},  \\
T_{ii}=P_{\phi}=\frac{\dot{\phi}^2}{2} - V(\phi)- \frac{(\nabla
\phi)^2}{6a^2}\, .
\end{eqnarray}
Note that, if
the gradient term were dominant, we would obtain
$P_\phi=-\frac{\rho_\phi}{3}$, not enough to drive inflation. 
We can now split the inflaton field in 
$$
\phi(t)=\phi_{0}(t)+\delta\phi({\bf x},t)\, ,
$$
where $\phi_{0}$ is the `classical' (infinite wavelength) field, that is 
the expectation value of the inflaton field 
on the initial isotropic and
homogeneous state, while $\delta\phi({\bf x},t)$ represents the quantum
fluctuations around $\phi_{0}$.
In this section, we shall be concerned only with the evolution of the
classical field $\phi_0$. The next section will be devoted to the
crucial issue of the evolution of quantum perturbations during inflation.
This separation is justified by the fact that quantum fluctuations are much
smaller than the classical value and therefore negligible when looking at the 
classical evolution. Not to be overwhelmed by the notation, we shall
indicate the classical value of the inflaton field by $\phi$ from now on.
The energy momentum tensor becomes
\begin{eqnarray}
T_{00}=\rho_{\phi}=\frac{\dot{\phi}^2}{2} + V(\phi)\\
T_{ii}=P_{\phi}=\frac{\dot{\phi}^2}{2} - V(\phi).
\end{eqnarray}
If
$$
V(\phi) \gg \dot{\phi}^2
$$
we obtain the following condition
$$
P_\phi\simeq -\rho_\phi\, .
$$
From this simple calculation, 
we realize that a scalar field whose energy is dominant
in the universe and whose potential energy  
dominates over the kinetic term gives inflation. Inflation
is driven by the vacuum energy of the inflaton field.

\subsection{Slow-roll conditions}

Let us now quantify  better under which circumstances a scalar field
may give rise to a period of inflation. 
The equation of motion of the field is
 \begin{equation}
 \ddot{\phi}+3H\dot{\phi}+V'(\phi)=0\, .
\label{poi}
 \end{equation}
If we require that $\dot{\phi}^2\ll V(\phi)$, the scalar field 
is slowly rolling down
its potential. This is the reason why such  a period is called {\it slow-roll}.
We may also expect that since the potential is flat, 
$\ddot{\phi}$ is negligible as well. We
shall assume that this is true and we shall quantify this condition soon.
The FRW equation (\ref{fr1}) becomes
\begin{equation}
H^2\simeq \frac{8\pi G}{3}\,V(\phi),
\end{equation}
where we have assumed that the inflaton field dominates the
energy density of the universe.
The new equation of motion becomes
\begin{equation}
 3H\dot{\phi}=-V'(\phi)
\label{friction}
\end{equation}
which gives $\dot{\phi}$ as a function of $V'(\phi)$.
Using Eq. (\ref{friction}) slow-roll  conditions then require
$$
\dot\phi^2 \ll  V(\phi)   \\  \Longrightarrow  \\  \frac{(V')^2}{V} \ll
H^2 \label{slowroll1}
$$
and
$$
\ddot{\phi} \ll 3H\dot{\phi} \\  \Longrightarrow  \\  V'' \ll H^2.  
\label{slowroll2}
$$
It is now useful to define the  slow-roll 
parameters $\epsilon$ and $\eta$ in the following way
\begin{center}
\begin{tabular}{|p{13 cm}|}
\hline
%\\
\begin{eqnarray}
\epsilon&=&-\frac{\dot{H}}{H^2}=4\pi G\frac{\dot{\phi}^2}{H^2}
=\frac{1}{16\pi
G}\left(\frac{V'}{V}\right)^2, \label{epsilon slow roll}\nonumber\\
\eta&=&\frac{1}{8\pi G} \left(\frac{V''}{V}\right)=\frac{1}{3}
\frac{V''}{H^2},\nonumber\\
\delta&=&\eta-\epsilon=-\frac{\ddot{\phi}}{H \dot{\phi}}\, 
.\nonumber
\end{eqnarray}
\\
\hline
\end{tabular}
\end{center}
It might be useful to have the same parameters expressed in terms of
conformal time
\begin{center}
\begin{tabular}{|p{13 cm}|}
\hline
%\\
\begin{eqnarray}
\epsilon&=&1-\frac{\H^\prime}{\H^2}=4\pi G\frac{\phi{^\prime}^2}{\H^2}
\label{epsilon slow roll conf}\nonumber\\
\delta&=&\eta-\epsilon=1-\frac{\phi^{\prime\prime}}{\H \phi^\prime}
\, .\nonumber
\end{eqnarray}
%\\
\\
\hline
\end{tabular}
\end{center}
The parameter $\epsilon$ quantifies how much the
Hubble rate $H$ changes with time during inflation. Notice that, since
$$
\frac{\ddot a}{a}=\dot H+H^2=\left(1-\epsilon\right)H^2,
$$
inflation can be attained only if $\epsilon<1$:
\begin{center}
\begin{tabular}{|p{13 cm}|}
\hline
%\\
$$
{\rm INFLATION}~~~\Longleftrightarrow ~~~\epsilon <1.
$$
%\\
\\
\hline
\end{tabular}
\end{center}
As soon as this condition fails, inflation ends. In general, slow-roll 
inflation
is attained if $\epsilon\ll 1$ and $|\eta|\ll 1$. During inflation
the  slow-roll parameters $\epsilon$ and $\eta$ can be considered
to be approximately constant  since the potential $V(\phi)$
is very flat.

\vskip 0.2cm

{\it \underline{Comment}:} In the following, we shall work at {\it
first-order} perturbation in the slow-roll parameters, that is we shall
take only the first power of them. Since, using their definition, it is
easy to see that $\dot\epsilon,\dot\eta={\cal O}
\left(\epsilon^2,\eta^2\right)$, this amounts to saying that we shall
treat the slow-roll parameters as constant in time.
\vskip 0.2cm

Within these approximations, it is easy to compute the number of
e-foldings between the beginning and the end of inflation.
If we indicate by $\phi_i$ and  $\phi_f$ the values of the inflaton
field at the beginning and at the end of inflation, respectively,
we find that the {\it total} number of e-foldings is
\begin{eqnarray}
N&\equiv&\int_{t_i}^{t_f}\,H\,dt\nonumber\\
&\simeq& H\int^{\phi_f}_{\phi_i}
\frac{d\phi}{\dot{\phi}}\nonumber\\
&\simeq&-3 H^2\int^{\phi_f}_{\phi_i}
\frac{d\phi}{V'}\nonumber\\
&\simeq& -8\pi G \int^{\phi_f}_{\phi_i} \frac{V}{V'}\,d\phi\, .
 \end{eqnarray}

We may also compute the number of e-foldings $\Delta N$ which are left to go
to the end of inflation

\begin{equation}
\label{togo}
\Delta N\simeq  8\pi G \int^{\phi_{\Delta N}}_{\phi_f}
\frac{V}{V'}\,d\phi,
\end{equation}
where $\phi_{\Delta N}$ is the value of the inflaton field
when there are $\Delta N$ e-foldings to the end of inflation.

{\it 1. \underline{Comment}:} According to the criterion 
given in Subsection 2.4, a given scale
length $\lambda=a/k$ leaves the horizon when $k=aH_k$
where
$H_k$ is the 
the value of the Hubble rate at that time. One can  easily compute
the rate of change of $H^2_k$ as a function of $k$
\begin{equation}
\frac{d {\rm ln} \,H_k^2}{d {\rm ln} \,k}=
\left(\frac{d {\rm ln} \,H_k^2}{dt}\right)\left(\frac{dt}{d {\rm ln} \,a}
\right)\left( \frac{d {\rm ln} \,a}{d {\rm ln} \,k}\right)=
2\frac{\dot H}{H}
\times \frac{1}{H}\times 
1=2\frac{\dot H}{H^2}=-2\epsilon.
\label{z}
\end{equation}

{\it 2. \underline{Comment}:} Take a  given physical scale $\lambda$ today 
which crossed the horizon scale during inflation. This happened when
$$
\lambda\left(\frac{a_f}{a_0}\right)e^{-\Delta N_\lambda}=\lambda
\left(\frac{T_0}{T_f}\right)e^{-\Delta N_\lambda}=H_I^{-1}
$$
where $\Delta N_\lambda$ indicates the number of e-foldings from the time the
scale crossed the horizon during inflation and the end of inflation.
This relation gives a way to determine the number of e-foldings 
to the end of inflation corresponding to a given scale
$$
\Delta N_\lambda\simeq 65 +{\rm ln}\left(\frac{\lambda}{3000\,\,{\rm Mpc}}
\right)+2\,{\rm ln}\left(\frac{V^{1/4}}{10^{14}\,\,{\rm GeV}}
\right)-{\rm ln}\left(\frac{T_f}{10^{10}\,\,{\rm GeV}}
\right).
$$
Scales relevant  for the CMB anisotropies 
correspond  to $\Delta N\sim $60.

Inflation ended when the potential energy associated with the inflaton
field became smaller than the kinetic energy of the field.  By that
time, any pre-inflation entropy in the universe had been inflated
away, and the energy of the universe was entirely in the form of
coherent oscillations of the inflaton condensate around the minimum of
its potential.  The universe may be said to be frozen after the end of
inflation. We know that somehow the low-entropy cold universe
dominated by the energy of coherent motion of the $\phi$ field must be
transformed into a high-entropy hot universe dominated by
radiation. The process by which the energy of the inflaton field is
transferred from the inflaton field to radiation has been dubbed
{\it reheating}. In the theory of reheating, 
the simplest way to envisage this process is if the co-moving energy
density in the zero mode of the inflaton decays into normal particles,
which then scatter and thermalize to form a thermal background.  It is
usually assumed that the decay width of this process is the same as
the decay width of a free inflaton field.

Of particular interest is a quantity usually known 
as the reheat temperature,
denoted as $T_{RH}$\footnote{So far, we have indicated it by
$T_f$.}. The reheat temperature is calculated by assuming 
an instantaneous conversion of the energy density in the inflaton 
field into radiation when the decay width of the inflaton energy,
$\Gamma_\phi$, is equal to $H$, the expansion rate of the universe. 

The reheat temperature is calculated quite easily.   After inflation
the inflaton field executes coherent oscillations about the minimum
of the potential.  Averaged over several oscillations, the coherent
oscillation energy density redshifts as matter: $\rho_\phi \propto
a^{-3}$, where $a$ is the Robertson--Walker scale factor.  If we
denote as $\rho_I$ and $a_I$ the total inflaton energy density 
and the scale factor at the initiation of coherent oscillations,
then the Hubble expansion rate as a function of $a$ is 
\begin{equation}
H^2(a) = \frac{8\pi}{3}\frac{\rho_I}{\mpl^2}
	\left( \frac{a_I}{a} \right)^3.
\end{equation}
Equating $H(a)$ and $\Gamma_\phi$ leads to an expression for $a_I/a$.
Now if we assume that all available coherent energy density is
instantaneously converted into radiation at this value of $a_I/a$, we
can find the reheat temperature by setting the coherent energy
density, $\rho_\phi=\rho_I(a_I/a)^3$, equal to the radiation energy
density, $\rho_R=(\pi^2/30)g_*T_{RH}^4$, where $g_*$ is the effective
number of relativistic degrees of freedom at temperature $T_{RH}$.
The result is
\begin{equation}
\label{eq:TRH}
T_{RH} = \left( \frac{90}{8\pi^3g_*} \right)^{1/4}
	\sqrt{ \Gamma_\phi \mpl } \
       = 0.2 \left(\frac{200}{g_*}\right)^{1/4}
      \sqrt{ \Gamma_\phi \mpl } \ .
\end{equation}

\section{Inflation and the cosmological perturbations}
In order for structure formation to occur via gravitational
instability, there must have been small pre-existing fluctuations on
physical length scales when they crossed the Hubble radius in the 
radiation-dominated 
and matter-dominated 
 eras \cite{llbook}.  These fluctuations are given by inflation.
Indeed, In an exponentially expanding universe the  wavelenghts of all 
vacuum fluctuations of the inflaton field $\phi$ grow exponentially in the
expanding universe. When the wavelength of any particular fluctuation
becomes greater than $H^{-1}$, this fluctuation stops propagating, and 
its amplitude freezes at some non-zero value $\delta\phi$ because of the 
large
friction term $3H\dot\phi$  the equation of motion of the
field $\phi$. The amplitude of this fluctuation then remains 
almost unchanged for a very long time, whereas its wavelength grows 
exponentially to cosmological scales.

Once inflation has ended,
however, the Hubble radius increases faster than the scale factor, so
the fluctuations eventually re-enter the Hubble radius during the
radiation- or matter-dominated eras and provide the necessary seeds.

In summary, these are the key ingredients for understanding 
the observed structures in the universe within the
inflationary scenario:

\begin{itemize}

\item Quantum fluctuations of the inflaton field 
are excited during inflation and stretched to cosmological scales. At the
same time, being the inflaton fluctuations connected
to the metric perturbations through  Einstein's equations, 
ripples on the metric are also excited and stretched to cosmological 
scales.

\item Gravity acts a   messenger since it communicates 
the small seed perturbations to photons and baryons once a given wavelength becomes smaller
than the horizon scale after inflation.

\end{itemize}

Let us now see how quantum fluctuations are generated during inflation.
we shall proceed by steps. First, we shall consider the simplest
problem of studying the quantum fluctuations of a generic scalar field
during inflation: we shall
learn how perturbations evolve as a function of time and compute their
spectrum. Then---since a satisfactory description of the generation of
quantum fluctuations has to take both the inflaton and the metric
perturbations into account--- we shall study the system composed by 
quantum
fluctuations of the inflaton field  and  quantum fluctuations
of the metric.

\section{Quantum fluctuations of a generic massless scalar field during 
inflation}

Let us first see how the
fluctuations  of a generic scalar field $\chi$, which is {\it not}
the inflaton field, behave during inflation. To warm up we first
consider a de Sitter epoch during which the Hubble rate is
constant.

\subsection{Quantum fluctuations of a generic massless scalar field during 
a de Sitter stage}

We assume this field to be massless. The massive case will be analysed in 
the
next subsection.

Expanding the scalar field $\chi$ in Fourier modes
$$
\delta\chi({\bf x},t)=\int\,\frac{d^3{\bf k}}{(2\pi)^{3/2}}\,e^{i{\bf k}
\cdot{\bf x}}\,
\delta\chi_{{\bf k}}(t),
$$
we can write the equation for the
fluctuations as 
\be
\delta\ddot{\chi}_{\bf  k}
+3H\,\delta\dot{\chi}_{\bf k}+
\frac{k^2}{a^2}\,\delta\chi_{\bf k}=0\, .
\label{quantum}
\end{equation}
Let us study the qualitative behaviour of the solution to Eq. (\ref{quantum}).

\begin{itemize}

\item For wavelengths within the horizon, $\lambda\ll H^{-1}$, the 
corresponding wave-number satisfies the relation $k\gg a\,H$. In this
regime, we can neglect the friction term $3H\,\delta\dot{\chi}_{\bf k}$
and 
Eq. (\ref{quantum}) reduces to
\be
\delta\ddot{\chi}_{\bf k}+
\frac{k^2}{a^2}\,\delta\chi_{\bf k}=0,
\end{equation}
which is basically  the equation of motion of an harmonic oscillator.
Of course, the frequency term $k^2/a^2$ depends upon time because
the scale factor $a$ grows exponentially. On the qualitative level,
however, one expects that when the wavelength of the fluctuation is within
the horizon, the fluctuation oscillates.

\item For wavelengths above the horizon, $\lambda\gg H^{-1}$, the 
corresponding wave-number satisfies the relation $k\ll aH$
and the term $k^2/a^2$ can be safely neglected. Equation (\ref{quantum}) reduces to
\be
\delta\ddot{\chi}_{\bf  k}
+3H\,\delta\dot{\chi}_{\bf k}=0,
\end{equation}
which tells us that on superhorizon scales $\delta\chi_{\bf k}$ remains 
constant.

\end{itemize}

We have therefore the following picture: take a given fluctuation
whose initial wavelength $\lambda\sim a/k$ is within the horizon. The
fluctuations oscillate till the wavelength becomes of the order
of the horizon scale. When the wavelength crosses the horizon, the
fluctuation ceases to oscillate and gets frozen in.

Let us now study the evolution of the fluctuation in a more quantitative
way. To do so, we perform the following redefinition
$$
\delta\chi_{\bf k}=\frac{\delta\sigma_{\bf k}}{a}
$$
and we work in conformal time $d\tau=dt/a$. For the time
being, we solve the problem for a pure de Sitter expansion
and we take the scale 
factor
 exponentially growing as $a\sim e^{Ht}$; 
the corresponding
conformal factor reads (after choosing properly the integration constants) 
$$
a(\tau)=-\frac{1}{H\tau}\,\,\,\,(\tau<0).
$$
In the following we shall also solve the problem in the
case of quasi de Sitter expansion.
The beginning of inflation coincides with some initial time $\tau_i\ll 0$.
We find that Eq. (\ref{quantum})
becomes
\be
\delta\sigma^{\prime\prime}_{\bf k}+
\left(k^2-\frac{a^{\prime\prime}}{a}\right)\delta\sigma_{\bf k}=0.
\label{qq}
\end{equation}
We obtain an equation which is very `close' to the equation for a 
Klein--Gordon scalar field in flat space-time, the only difference being
a negative time-dependent mass term $-a^{\prime\prime}/a=-2/\tau^2$.
Equation (\ref{qq}) can be obtained from an action of the type
\be
\delta S_{\bf k}=\int\,d\tau\,\left[\frac{1}{2}
\delta\sigma^{\prime 2}_{\bf k}-\frac{1}{2}
\left(k^2-\frac{a^{\prime\prime}}{a}\right)\delta\sigma^2_{\bf k}
\right],
\label{action}
\end{equation}
which is the canonical action for a simple harmonic oscillator with
canonical commutation relations $\delta\sigma^*_{\bf 
k}\delta\sigma^\prime_{\bf k}
-\delta\sigma_{\bf k}\delta\sigma^{*\prime}_{\bf k}=-i$.

Let us study the behaviour of this equation on subhorizon and superhorizon
scales. Since 
$$
\frac{k}{aH}=-k\,\tau\,,
$$
on subhorizon scales $k^2\gg a^{\prime\prime}/a$   Equation (\ref{qq})
reduces to 
$$
\delta\sigma^{\prime\prime}_{\bf k}+
k^2\,\delta\sigma_{\bf k}=0\,,
$$
whose solution is a plane wave
\be
\delta\sigma_{\bf k}=\frac{e^{-ik\tau}}{\sqrt{2k}}\,\,\,\,(k\gg aH)\,.
\label{q1}
\end{equation}
We find again that  fluctuations with wavelength within the horizon
oscillate exactly like in flat space-time. 
This does not come as a surprise. In the 
ultraviolet regime, that is for wavelengths much smaller than the horizon
scale, one expects that approximating the space-time as flat
is a good approximation.

On superhorizon scales, 
$k^2\ll a^{\prime\prime}/a$ Equation (\ref{qq})
reduces to 
$$
\delta\sigma^{\prime\prime}_{\bf k}-
\frac{a^{\prime\prime}}{a}\delta\sigma_{\bf k}=0,
$$
which is satisfied by 
\be
\delta\sigma_{\bf k}=B(k)\,a \,\,\,\,(k\ll aH)\,
\label{x2}
\end{equation}
where $B(k)$ is a constant of integration. Roughly matching  the (absolute
values of the) solutions
$(\ref{q1})$ and $(\ref{x2})$ at $k=aH$ ($-k\tau=1$), we can determine the
(absolute value of the) constant $B(k)$
$$
\left|B(k)\right|a=\frac{1}{\sqrt{2k}}\Longrightarrow
\left|B(k)\right|=\frac{1}{a\sqrt{2k}}=\frac{H}{\sqrt{2k^3}}.
$$
Going back to the original variable 
$\delta\chi_{\bf k}$, we obtain that the quantum fluctuation of the 
$\chi$
field on superhorizon scales is constant and approximately
equal to
\begin{center}
\begin{tabular}{|p{13 cm}|}
\hline
%\\
$$
\left|\delta\chi_{\bf k}\right|\simeq \frac{H}{\sqrt{2k^3}}\,\,\,\,
({\rm ON}\,\,{\rm SUPERHORIZON}\,\,{\rm SCALES})
$$
%\\
\\
\hline
\end{tabular}
\end{center}
In fact we can do much better, since Eq. (\ref{qq}) has an {\it exact} 
solution:
\be
\label{sigma}
\delta\sigma_{\bf k}=\frac{e^{-ik\tau}}{\sqrt{2k}}\left(
1+\frac{i}{k\tau}\right).
\end{equation}
This solution reproduces all that we have found by qualitative arguments
in the two extreme regimes $k\ll aH$ and $k\gg aH$. We have
performed the matching procedure  to show that the latter can be
very useful to determine the behaviour of the solution on superhorizon scales
when the exact solution is not known.

\subsection{The power spectrum}

Let us define now the power spectrum, a useful quantity to
characterize the properties of the perturbations. 
For a generic quantity $g({\bf x},t)$, which can expanded in
Fourier space as
$$
g({\bf x},t)=\int\,\frac{d^3{\bf k}}{(2\pi)^{3/2}}\,e^{i{\bf k}
\cdot{\bf x}}\,
g_{{\bf k}}(t),
$$
the power spectrum can be defined as
\be
\langle 0|g^{*}_{{\bf k}_1}g_{{\bf k}_2}|0\rangle
\equiv\delta^{(3)}\left({\bf k}_1-{\bf k}_2\right)\,\frac{2\pi^2}{k^3}\,
{\cal P}_{g}(k),
\end{equation}
where $\left|0\right.\rangle$ is the vacuum quantum state of the system. This
definition leads to the usual relation
\be
\langle 0|g^2({\bf x},t)|0\rangle=\int\,\frac{dk}{k}\,
{\cal P}_{g}(k).
\end{equation}

\subsection{Quantum fluctuations of a generic scalar field in a quasi de 
Sitter stage}

So far, we have computed the time evolution and the spectrum of the 
quantum fluctuations of a generic scalar field $\chi$ 
supposing that the 
scale factor evolves
like in a pure de Sitter expansion, $a(\tau)=-1/(H\tau)$. However, 
during
inflation the Hubble rate is not exactly constant, but changes with time
as $\dot H=-\epsilon\,H^2$ (quasi de Sitter expansion).
In this subsection, we shall solve for the perturbations
in a quasi de Sitter expansion. Using the definition of the
conformal time, one can show that the scale factor for small values
of $\epsilon$
becomes
$$
a(\tau)=-\frac{1}{H}\frac{1}{\tau(1-\epsilon)}.
$$
The fluctuation mass-squared mass term is
$$
M^2(\tau)=m_\chi^2a^2-\frac{a^{\prime\prime}}{a},
$$
where
\begin{eqnarray}
\label{ap}
\frac{a^{\prime\prime}}{a}&=&a^2\left(\frac{\ddot{a}}{a}+H^2\right)=
a^2\left(\dot H+2\,H^2\right)\nonumber\\
&=&a^2\left(2-\epsilon\right)H^2=
\frac{\left(2-\epsilon\right)}{\tau^2\left(1-\epsilon\right)^2}\nonumber\\
&\simeq& \frac{1}{\tau^2}\left(2+3\epsilon\right).
\end{eqnarray}
Armed with these results, we may  compute the
variance of the perturbations of the generic $\chi$ field
\begin{eqnarray}
\langle 0|\left(\delta\chi({\bf x},t)\right)^2|0\rangle&=&
\int\,\frac{d^3k}{(2\pi)^3}\,\left|\delta\chi_{\bf k}\right|^2\nonumber\\
&=&\int\,\frac{dk}{k}\,\frac{k^3}{2\pi^2}
\,\left|\delta\chi_{\bf k}\right|^2\nonumber\\
&=& \int\,\frac{dk}{k}\,{\cal P}_{\delta\chi}(k),
\end{eqnarray}
which defines the power spectrum of the
fluctuations of the scalar field $\chi$
\be
{\cal P}_{\delta\chi}(k)\equiv\frac{k^3}{2\pi^2}
\,\left|\delta\chi_{\bf k}\right|^2.
\label{spectrum}
\end{equation}
Since we have seen that fluctuations are (nearly) 
frozen in on superhorizon scales,
a way of characterizing the perturbations is to compute
the spectrum on scales larger than the horizon. For a massive scalar 
field, we obtain
\be
{\cal P}_{\delta\chi}(k)=\left(\frac{H}{2\pi}\right)^2
\left(\frac{k}{aH}\right)^{3-2\nu_\chi},
\label{fff}
\end{equation}
where, taking $m_\chi^2/H^2=3\eta_\chi$ and expanding for small values
of $\epsilon$ and $\eta$,  
\be
\label{vv}
\nu_\chi\simeq \frac{3}{2}+\epsilon-\eta_\chi.
\end{equation}
We may also define the {\it spectral index} $n_{\delta\chi}$ 
of the fluctuations
as
\begin{center}
\begin{tabular}{|p{13 cm}|}
\hline
%\\
$$
n_{\delta\chi}-1=
\frac{d {\rm ln} \,{\cal P}_{\delta\phi}}{d {\rm ln} \,k}=3-2\nu_\chi=
2\eta_\chi-2\epsilon.
$$
\\
\hline
\end{tabular}
\end{center}
The power spectrum of fluctuations of the scalar
field $\chi$  is therefore
{\it nearly flat}, that is is nearly independent of the wavelength
$\lambda=\pi/k$: the amplitude of the 
fluctuation on superhorizon scales does almost not  depend upon the 
time at which the fluctuation crosses the horizon and becomes frozen
in. The small tilt of the power spectrum arises from the fact that
the scalar field $\chi$ is massive and because 
during inflation the Hubble rate is not exactly constant, but
nearly constant, where `nearly' is quantified by the slow-roll
parameters $\epsilon$. Adopting the  traditional 
terminology,
we may say that the spectrum of perturbations is blue if 
$n_{\delta\chi}>1$
(more power in the ultraviolet)
and red if $n_{\delta\chi}<1$ (more power in the infrared).
The power spectrum of the perturbations
of a generic scalar field $\chi$ generated during a period
of slow-roll inflation may be either blue or red. This
depends upon the relative magnitude between $\eta_\chi$ and $\epsilon$.

\vskip 0.2cm

{\it \underline{Comment}:} We might have computed the
spectral index of the spectrum ${\cal P}_{\delta\chi}(k)$ by first
solving the equation for the perturbations of the field $\chi$
in a di Sitter stage, with $H=$ constant and therefore $\epsilon=0$, and
then taking into account the time evolution of the Hubble rate
introducing  the subscript in $H_k$ whose time variation is determined 
by  Eq. (\ref{z}). Correspondingly, $H_k$ 
is the value of the Hubble rate  when a given wavelength $\sim 
k^{-1}$ crosses
the horizon (from that point on the fluctuation remains
frozen in). The power spectrum in such an approach would read

\be
{\cal P}_{\delta\chi}(k)=\left(\frac{H_k}{2\pi}\right)^2
\left(\frac{k}{aH}\right)^{3-2\nu_\chi}
\label{bb}
\end{equation}
with $3-2\nu_\chi\simeq \eta_\chi$. Using Eq. (\ref{z}), one finds

$$
n_{\delta\chi}-1=
\frac{d {\rm ln} \,{\cal P}_{\delta\phi}}{d {\rm ln} \,k}=
\frac{d {\rm ln} \,H_k^2}{d {\rm ln} \, k}+3-2\nu_\chi=
2\eta_\chi-2\epsilon
$$
which 
reproduces our previous findings.

\vskip 0.2cm   

{\it \underline{Comment}:} Since on superhorizon scales 

$$
\delta\chi_{\bf k}\simeq \frac{H}{\sqrt{2k^3}}
\left(\frac{k}{aH}\right)^{\eta_\chi-\epsilon}\simeq
\frac{H}{\sqrt{2k^3}}\left[1+\left(\eta_\chi-\epsilon\right)
{\rm ln}\,\left(\frac{k}{aH}\right)\right],
$$
we discover that 

\begin{equation}
\label{e}
\left|\delta\dot{\chi}_{\bf k}\right|\simeq \left|
H\left(\eta_\chi-\epsilon\right)
\,\delta\chi_{\bf k}\right|\ll \left|H\,\delta\chi_{\bf k}\right|,
\end{equation}
that is,  on superhorizon scales the time variation of the 
perturbations can be safely neglected.

\section{Quantum fluctuations during inflation}

As we have mentioned in the previous section, the linear theory of the
cosmological perturbations represents a cornerstone of modern cosmology
and is used to describe the formation and evolution of structures
in the universe as well as the anisotropies of the CMB. The seeds
for these inhomogeneities were generated during inflation and
stretched over astronomical scales because of the rapid superluminal
expansion of the universe during the (quasi) de Sitter epoch.

In the previous section we have already seen
that pertubations of a generic scalar field $\chi$ are generated
during a (quasi) de Sitter expansion. The inflaton
field is a scalar field and, as such, we conclude that 
inflaton fluctuations will be generated as well. However, 
the inflaton is special from the point of view
of perturbations. The reason is very simple. By assumption, the
inflaton field dominates the energy density of the universe during
inflation. Any perturbation in the inflaton field means a perturbation
of the stress energy momentum tensor

$$
\delta\phi\Longrightarrow \delta T_{\mu\nu}.
$$
A perturbation in the stress energy momentum tensor implies,
through Einstein's equations of motion, a perturbation of the metric

$$
\delta T_{\mu\nu}\Longrightarrow \left[
\delta R_{\mu\nu}-\frac{1}{2}\delta\left(g_{\mu\nu}R\right)\right]
=8\pi G\delta T_{\mu\nu}\Longrightarrow \delta g_{\mu\nu}.
$$
On the other hand, a pertubation of the metric induces a back-reaction
on the evolution of the inflaton perturbation through the 
perturbed Klein--Gordon equation of the inflaton field

$$
\delta g_{\mu\nu}\Longrightarrow \delta\left(\partial_\mu\partial^\mu\phi+
\frac{\partial V}{\partial\phi}\right)=0\Longrightarrow\delta\phi.
$$
This logic chain makes us conclude that the perturbations of the
inflaton field and of the metric are tightly coupled to each other
and have to be studied together

\begin{center}
\begin{tabular}{|p{13.0 cm}|}
\hline
%\\
$$
\delta\phi\Longleftrightarrow\delta g_{\mu\nu}\, .
$$
%\\
\\
\hline
\end{tabular}
\end{center}
As we shall see shortly, this
relation is stronger than one might think because of the issue
of gauge invariance.

Before launching ourselves into the problem of finding the evolution
of the quantum perturbations of  the inflaton field when they are coupled 
to gravity, 
let us give  a heuristic 
explanation of why we expect that during inflation such fluctuations are
indeed present.

If we take Eq. (\ref{nabla}) and split the  inflaton field as
its classical value $\phi_0$ plus the quantum flucutation $\delta\phi$,  
$\phi({\bf x},t)=\phi_{0}(t)+\delta\phi({\bf x},t)$, the quantum perturbation
$\delta\phi$ satisfies the equation of motion
\begin{equation}
\delta\ddot{\phi}+3H\,\delta{\dot\phi}-\frac{\nabla^2\delta\phi}{a^2}
+V''\,
\delta\phi=0.
\label{aa}
\end{equation}
Differentiating  Eq. (\ref{poi}) wrt time and taking $H$ constant
(de Sitter expansion)  we find
\begin{equation}
({\phi}_0)^{\cdot\cdot\cdot}+3H\ddot{\phi}_0 +V''\,\dot{\phi}_0=0.
\end{equation}
Let us consider for simplicity the limit ${\bf k}^2/a^2\ll 1$ 
and let us disregard
the gradient term. Under this condition we see that
$\dot{\phi}_0$ and $\delta\phi$ solve the same equation. The solutions
have therefore to be related to each other by a constant of proportionality
which depends upon time

\be
\label{old}
\delta\phi=-\dot{\phi}_0\,\delta t({\bf x}).
\end{equation}

This tells us that $\phi({\bf x},t)$ will have the form
$$
\phi({\bf x},t)=\phi_0\left({\bf x},t-\delta t({\bf x})\right).
$$

This equation indicates that the inflaton field does not acquire
the same value at a given time $t$ in all the space. On the contrary,
when the inflaton field is rolling down its potential, it acquires
different values from one spatial point ${\bf x}$ to the next. The inflaton
field is not homogeneous and fluctuations are present. These fluctuations, 
in turn, will induce fluctuations in the metric.

\subsection{The metric fluctuations}

The mathematical tool to describe the linear evolution
of the cosmological perturbations is obtained  by 
perturbing at  the first order the 
FRW metric $g^{(0)}_{\mu\nu}$, see Eq. (\ref{metric})

\begin{equation}
g_{\mu\nu}\quad = \quad  g^{(0)}_{\mu\nu}(t) \,+\, 
g_{\mu\nu}(\mathbf{x},t)\,; \qquad  g_{\mu\nu} \,\ll
\,g^{(0)}_{\mu\nu}\,.
\end{equation}
The metric perturbations can be decomposed according to their spin
with respect to a local rotation of the spatial
coordinates on hypersurfaces of constant time. This leads to

\begin{itemize}

\item {\it scalar perturbations}

\item {\it vector perturbations}

\item{\it tensor perturbations}

\end{itemize}

Tensor perturbations or gravitational waves 
have spin 2 and are the true degrees of 
freedom of the gravitational fields  in the sense that they can
exist even in the vacuum. Vector perturbations are spin 1 modes arising from
rotational velocity fields and are also called vorticity modes. Finally,
scalar perturbations have spin 0. 

Let us do  a simple exercise to count how many scalar degrees of freedom
are present. Take a space-time of dimensions $D=n+1$, of which $n$
coordinates are spatial coordinates. The symmetric metric tensor $g_{\mu\nu}$
has $\frac{1}{2}(n+2)(n+1)$ degrees of freedom. We can perform $(n+1)$
coordinate transformations in order to eliminate $(n+1)$ degrees of freedom,
this leaves us with $\frac{1}{2}n(n+1)$ degrees of freedom.
These $\frac{1}{2}n(n+1)$ degrees of freedom contain scalar, vector
and tensor modes. According to Helmholtz's theorem we can always decompose
a vector $u_i$ $(i=1,\cdots,n)$ as $u_i=\partial_i v +v_i$, where
$v$ is a scalar (usually called potential flow) which is curl-free,
$v_{[i,j]}=0$, 
and $v_i$ is a real vector (usually called vorticity) which is divergence-free,
$\nabla\cdot v=0$. This means that the real vector (vorticity) modes
are $(n-1)$. Furthermore, a generic traceless tensor $\Pi_{ij}$
can always be decomposed as $\Pi_{ij} =\Pi^S_{ij}+\Pi_{ij}^V+
\Pi_{ij}^T$, where $\Pi^S_{ij}=\left(-\frac{k_i k_j}{k^2}+
\frac{1}{3}\delta_{ij}\right)\Pi$, $\Pi^V_{ij}=(-i/2k)\left(k_i\Pi_j
+k_j\Pi_i\right)$ $(k_i\Pi_i=0)$  and $k_i\Pi^T_{ij}=0$. This means that the
true symmetric, traceless and transverse tensor degreees of freedom
are $\frac{1}{2}(n-2)(n+1)$. 

The number of
scalar degrees of freedom is therefore

$$
\frac{1}{2}n(n+1)-(n-1)-\frac{1}{2}(n-2)(n+1)=2,
$$
while the degrees of freedom  
of true vector modes are $(n-1)$ and the number of degrees of freedom of 
true
tensor modes (gravitational waves) is $\frac{1}{2}(n-2)(n+1)$. In four
dimensions $n=3$, meaning that one expects 2 scalar degrees of freedom,
2 vector degrees of freedom and 2 tensor degrees of freedom.
As we shall see, to the 2 scalar degrees of freedom from the 
metric, one has to add  another
one, the inflaton field perturbation $\delta\phi$. However, since
Einstein's equations will tell us that the two scalar degrees of freedom
from the metric are equal during inflation, we expect a total number
of scalar degrees of freedom equal to 2.

At the linear order, the scalar, vector, and tensor perturbations evolve
independently (they decouple) and it is therefore possible to analyse
them separately. Vector perturbations are not excited
during inflation because there are no rotational velocity fields
during the inflationary stage. we shall analyse the generation
of tensor modes (gravitational waves) in the following. For the
time being we want to focus on the scalar degrees of freedom of the metric.

Considering only the scalar degrees of freedom of the perturbed
metric, the most generic perturbed metric reads

\begin{equation}
g_{\mu\nu}\,=\, a^2 \left(
\begin{array}{c c}
- 1 \,-\, 2\,\Phi & \partial_i B \\
\partial_i B & \left( 1 \,-\, 2\,\psi\right)\delta_{ij} \,+\, D_{ij}
E \\
\end{array}
\right),
\end{equation}
while the line-element can be written as 
\begin{equation}
ds^2 \,=\, a^2 \big( ( - 1 - 2\,\Phi)d\tau^2 \,+\, 2 \,\partial_i B
\,d\tau\,dx^i \,+\, \left((1 - 2\,\psi)\delta_{ij} \,+\,
D_{ij}E\right) \,dx^i\,dx^j \big).
\end{equation}
Here $D_{ij}\,=\left(\partial_i
\partial_j \,-\, \frac{1}{3}\,\delta_{ij}\,\nabla^2\right)$.

\subsection{The issue of gauge invariance}

When studying the cosmological density perturbations, 
what we are interested in is following the evolution of a space-time which
is neither homogeneous nor isotropic.  This is done by following
the evolution of the differences between the actual space-time and a
well understood reference space-time.  So we shall consider small
perturbations away from the homogeneous, isotropic space-time.

The reference system in our case is the
spatially flat Friedmann--Robertson--Walker (FRW) space-time, with line
element $ds^2 = a^2(\tau) \left\{ d\tau^2 - \delta_{ij} dx^idx^j
\right\}$.
Now, the key issue is that general relativity is a gauge theory where
the gauge transformations are the generic coordinate transformations
from one local reference frame to another. 

When we compute the perturbation of a given quantity, this is defined
to be the difference between the value that this quantity assumes on the
real physical  space-time and the value it assumes on the unperturbed 
background. Nonetheless, to perform a comparison between these two
values, it is necessary to compute them at the same space-time point.
Since the two values   live   on two different geometries, it is necessary to
specify a map which allows one to link univocally the
same point on the two different space-times. This correspondence is called
a gauge choice and changing the map means performing a gauge transformation.

Fixing a gauge in general relativity implies choosing
a coordinate system. A choice of coordinates 
defines a {\it threading} of space-time into 
lines (corresponding to fixed spatial coordinates  ${\bf x}$) and a 
{\it slicing} into hypersurfaces (corresponding to fixed time $\tau$).
A choice of coordinates is  called a {\it gauge} and there is no unique
preferred gauge
\begin{center}
\begin{tabular}{|p{13.0 cm}|}
\hline
%\\
$$
{\rm GAUGE ~~CHOICE}~~~\Longleftrightarrow~~~{\rm SLICING ~~AND ~~THREADING}
$$
%\\
\\
\hline
\end{tabular}
\end{center}
From a more formal point of view, operating  an infinitesimal 
 gauge transformation on the coordinates
\begin{equation}
\label{gauge} \widetilde{x^\mu}\,=\, x^\mu \,+\, \delta x^\mu
\end{equation}
implies on a generic quantity $Q$ a transformation on its
perturbation
\begin{equation}
\widetilde{\deu Q}\,=\, \deu Q \,+\,\pounds_{\delta x} \,Q_0\,
\label{formal}
\end{equation}
where $Q_0$ is the value assumed by the quantity $Q$ on the background
and $\pounds_{\delta x}$ is the Lie-derivative of $Q$ along
the vector $\delta x^\mu$.

Decomposing in the
usual manner the vector $\delta x^\mu$ 
\begin{eqnarray}
 \delta x^0 \,&=&\, \xi^0(x^\mu)\,; \nonumber\\
\delta x^i \,&=&\, \partial^i \beta(x^\mu) \,+\, v^i(x^\mu) 
\,; \qquad \partial_i v^i
\,=\,0\,,
\end{eqnarray}
we can easily deduce the transformation
law of a scalar quantity $f$ (like the inflaton scalar field $\phi$
and energy density $\rho$). Instead of applying the formal definition
(\ref{formal}), we find the transformation law in an alternative (and 
more pedagogical) way.
We first write $\delta f(x)=f(x)-f_0(x)$,
where $f_0(x)$ is the background value. Under a gauge transformation
we have $\widetilde{\delta f}(\widetilde{x^\mu})=
\widetilde{f}(\widetilde{x^\mu})-\widetilde{f}_0(\widetilde{x^\mu})$.
Since $f$ is a scalar we can write $f(\widetilde{x^\mu})=f(x^\mu)$
(the value of the scalar function in a given physical point is
the same in all the coordinate system). On the other side, on the
unperturbed background hypersurface $\widetilde{f}_0=
f_0$. We have therefore

\begin{eqnarray}
\widetilde{\delta f}(\widetilde{x^\mu})&=&
\widetilde{f}(\widetilde{x^\mu})-\widetilde{f}_0(\widetilde{x^\mu})\nonumber\\
&=& f(x^\mu)-f_0(\widetilde{x^\mu})\nonumber\\
&=&f\left(\widetilde{x^\mu}\right)-f_0(\widetilde{x^\mu})\nonumber\\
&=&f(\widetilde{x^\mu})-\delta x^\mu\,\frac{\partial f}{\partial x^\mu}
(\widetilde{x})-f_0(\widetilde{x^\mu}),\nonumber\\
\end{eqnarray}
from which we finally deduce, being $f_0=f_0(x^0)$, 

\begin{center}
\begin{tabular}{|p{13 cm}|}
\hline
$$
\widetilde{\delta f}=\delta f-f^\prime\,\xi^0
$$
\\
\hline
\end{tabular}
\end{center}
For the spin-zero perturbations of  the metric, we can proceed analogously.
We use the following trick. Upon a coordinate transformation
$x^\mu\rightarrow \widetilde{x^\mu}=x^\mu+\delta x^\mu$, the line
element is left invariant, $ds^2=\widetilde{ds^2}$. This implies, for instance,
that $a^2(\widetilde{x^0})\left(1+\widetilde{\Phi}\right)\left(
d\widetilde{x^0}\right)^2=
a^2(x^0)\left(1+\Phi\right)(dx^0)^2$. Since 
$a^2(\widetilde{x^0})\simeq a^2(x^0)+2 a\, a^\prime\,\xi^0$ and
$d\widetilde{x^0}=\left(1+\xi^{0\prime}\right)dx^0+
\frac{\partial x^0}{\partial x^i}\,d x^i$, we obtain 
$1+2 \Phi=1+2\widetilde{\Phi}+2\H\xi^0+2\xi^{0\prime}$.
We now may introduce in detail 
some gauge-invariant quantities
which play  a major role in the computation of the
density perturbations. 
In the following we shall be interested only in the 
coordinate  transformations on constant time hypersurfaces
and therefore gauge invariance will be equivalent to independence of the 
slicing.

\subsection{The co-moving curvature perturbation}

The intrinsic spatial curvature on hypersurfaces on constant
conformal  time $\tau$ and for a flat universe is given by

$$
^{(3)}R=\frac{4}{a^2}\nabla^2\,\psi.
$$
The quantity $\psi$ is usually referred to as the {\it curvature perturbation}.
We have seen, however, that the curvature potential
$\psi$ is {\it not} gauge invariant, but is defined only on a 
given slicing.
Under a transformation on constant time hypersurfaces
$t\rightarrow t+\delta \tau$ (change of the slicing)

$$
\psi\rightarrow \psi+\H\,\delta\tau.
$$
We now consider the {\it co-moving slicing} which is defined to be the
slicing orthogonal to the worldlines of 
co-moving observers. The latter are    are free-falling
and the expansion defined by them is isotropic. In practice, what
this means is that there is no flux of energy measured by these
observers, that is $T_{0i}=0$. During inflation this means
that these observers measure $\delta\phi_{\rm com}=0$ since 
$T_{0i}$ goes like $\partial_i\delta\phi({\bf x},\tau)\phi^\prime(\tau)$.

Since $\delta\phi\rightarrow \delta\phi-\phi^\prime\delta\tau$ for a 
transformation on constant time hypersurfaces, this means that

$$
\delta\phi\rightarrow\delta\phi_{\rm com}=\delta\phi-\phi^\prime\,\delta\tau=0
\Longrightarrow \delta\tau=\frac{\delta\phi}{\phi^\prime},
$$
that is $\delta\tau=\frac{\delta\phi}{\phi^\prime}$ is the time-displacement
needed to go from a generic slicing with generic $\delta\phi$ to the
co-moving slicing where $\delta\phi_{\rm com}=0$.
At the same time the curvature perturbation $\psi$ transforms
into

$$
\psi\rightarrow\psi_{\rm com}= \psi+\H\,\delta\tau=\psi+
\H\frac{\delta\phi}{\phi^\prime}.
$$
The quantity

\begin{center}
\begin{tabular}{|p{13.0 cm}|}
\hline
%\\
$$
{\cal R}=\psi+
\H\frac{\delta\phi}{\phi^\prime}=\psi+H\frac{\delta\phi}{\dot\phi}
$$
%\\
\\
\hline
\end{tabular}
\end{center}
is the {\it co-moving curvature perturbation}. This quantity is gauge invariant
by construction and is related to the gauge-dependent
curvature perturbation $\psi$ on a generic slicing to the inflaton
perturbation $\delta\phi$ in that gauge. By construction, the meaning of
${\cal R}$ is that it represents the gravitational potential on 
co-moving hypersurfaces where $\delta\phi=0$ or the inflaton fluctuation
hypersurfaces where $\psi=0$:

$$
{\cal R}=\left.\psi\right|_{\delta\phi=0}=\left.H\frac{\delta\phi}{\dot\phi}
\right|_{\psi=0}.
$$
The power spectrum of the curvature perturbation may then be 
easily computed 

\begin{equation}
{\cal R}_{\bf k}=H\, \frac{\delta\phi_{\bf k}}{\dot\phi}.
\end{equation}
We may now compute the power spectrum of the co-moving curvature
perturbation on superhorizon scales

\begin{center}
\begin{tabular}{|p{13 cm}|}
\hline
%\\
$$
{\cal P}_{{\cal 
R}}(k)=\frac{1}{2\mpl^2\epsilon}\left(\frac{H}{2\pi}\right)^2
\left(\frac{k}{aH}\right)^{n_{{\cal R}}-1}\equiv A^2_{\cal R}
\left(\frac{k}{aH}\right)^{n_{{\cal R}}-1}
$$
\\
\hline
\end{tabular}
\end{center}
where we have defined the {\it spectral index} $n_{{\cal R}}$ of the co-moving
curvature perturbation
as
\begin{center}
\begin{tabular}{|p{13 cm}|}
\hline
%\\
$$
n_{{\cal R}}-1=
\frac{d {\rm ln} \,{\cal P}_{{\cal R}}}{d {\rm ln} \,k}=3-2\nu=
2\eta-6\epsilon.
$$
\\
\hline
\end{tabular}
\end{center}
\vskip 0.2cm
We conclude that inflation is responsible for the generation of 
adiabatic/curvature perturbations with an almost scale-independent
spectrum. To   compute the
spectral index of the spectrum ${\cal P}_{{\cal R}}(k)$ we have proceeded as follows:  first
solve the equation for the perturbation $\delta\phi_{\bf k}$
in a de Sitter stage, with $H=$ constant ($\epsilon=\eta=0$),
whose solution is Eq. (\ref{sigma}) and
then taking into account the time-evolution of the Hubble rate
and of $\phi$ introducing  the subscript in $H_k$ and $\dot{\phi}_k$. The 
time variation of the latter  is determined 
by 
\be
\frac{d {\rm ln} \,\dot{\phi}_k}{d {\rm ln} \,k}=
\left(\frac{d {\rm ln} \,\dot{\phi}_k}{dt}\right)
\left(\frac{dt}{d {\rm ln} \,a}
\right)\left(\frac{d {\rm ln} \,a}{d {\rm ln} \,k}\right)=
\frac{\ddot{\phi}_k}{\dot\phi_k}
\times \frac{1}{H}\times 
1=-\delta=\epsilon-\eta.
\label{zzz}
\end{equation}
Correspondingly, $\dot{\phi}_k$ 
is the value of the time derivative of the inflaton field
  when a given wavelength $\sim 
k^{-1}$ crosses
the horizon (from that point on the fluctuations remains
frozen in). The curvature 
perturbation in such an approach would read

$$
{\cal R}_{\bf k}\simeq \frac{H_k}{\dot\phi_k}\,\delta\phi_{\bf k}\simeq
\frac{1}{2\pi}\left(\frac{H_k^2}{\dot\phi_k}\right).
$$
Correspondingly

$$
n_{{\cal R}}-1=
\frac{d {\rm ln} \,{\cal P}_{{\cal R}}}{d {\rm ln} \,k}=
\frac{d {\rm ln} \,H_k^4}{d {\rm ln} \, k}-
\frac{d {\rm ln} \,\dot{\phi}_k^2}{d {\rm ln} \, k}=
-4\epsilon+(2\eta-2\epsilon)=2\eta-6\epsilon.
$$

During inflation the curvature
perturbation is generated on superhorizon scales with a spectrum which
is nearly scale invariant \cite{muk},  that is, is nearly independent of the wavelength
$\lambda=\pi/k$: the amplitude of the 
fluctuation on superhorizon scales does not (almost) depend upon the 
time at which the fluctuation crosses the horizon and becomes frozen
in. The small tilt of the power spectrum arises from the fact that
the inflaton field  is massive, giving rise to a non-vanishing $\eta$
 and because 
during inflation the Hubble rate is not exactly constant, but
nearly constant, where `nearly' is quantified by the slow-roll
parameters $\epsilon$. 

\vskip 0.2cm

{\it \underline{Comment}:} From what we have found so far, we may conclude that
on superhorizon scales the co-moving curvature perturbation ${\cal R}$
and the uniform-density gauge curvature $\zeta$ satisfy on superhorizon scales
the relation

$$
\dot{\cal R}_{\bf k}\simeq 0.
$$

\subsection{Gravitational waves}

Quantum fluctuations in the gravitational fields are generated 
in a similar fashion to that of the scalar perturbations 
discussed so far. A gravitational wave
may be viewed as a ripple of space-time in the FRW background metric
(\ref{metric}) and in general the linear tensor perturbations
may be written as 

$$
g_{\mu\nu}=a^2(\tau)\left[-d\tau^2+\left(\delta_{ij}+h_{ij}\right)
dx^i dx^j\right],
$$
where $\left|h_{ij}\right|\ll 1$. The tensor $h_{ij}$ has six degrees of 
freedom, but, as we studied in Subsection 7.1,
 the tensor perturbations are traceless, $\delta^{ij}h_{ij}=0$,
and transverse $\partial^i h_{ij}=0$ $(i=1,2,3)$. With these
four constraints, there remain two physical degrees of freedom, or 
polarizations, 
 which are
usually indicated 
$\lambda=+,\times$. More precisely, we can write

$$
h_{ij}=h_+\,e_{ij}^+ +h_\times\,e_{ij}^\times,
$$
where $e^+$ and $e^\times$ are the polarization tensors which 
have the following properties

$$
e_{ij}=e_{ji},~~~ k^i e_{ij}=0,~~~,e_{ii}=0,
$$
$$
e_{ij}(-{\bf k},\lambda)=e^*_{ij}({\bf k},\lambda),~~~
\sum_\lambda\,e^*_{ij}({\bf k},\lambda)e^{ij}({\bf k},\lambda)=4.
$$
Notice also that
the tensors $h_{ij}$ are gauge-invariant and therefore represent physical
degrees of freedom.

If the stress-energy momentum tensor is diagonal, as the one 
provided by the inflaton potential $T_{\mu\nu}=\partial_\mu\phi
\partial_\nu\phi-g_{\mu\nu}{\cal L}$, the tensor modes do not have any
source in their equation and their action can be written as 

$$
\frac{\mpl^2}{2}\,\int\,d^4x\,\sqrt{-g}\, \frac{1}{2}\partial_\sigma
h_{ij}\,\partial^\sigma h_{ij},
$$
that is the action of four independent massless scalar fields. The 
gauge-invariant tensor amplitude

$$
v_{\bf k}=a\mpl\frac{1}{\sqrt{2}}\, h_{\bf k},
$$
satisfies therefore the equation

$$
v_{\bf k}^{\prime\prime}+\left(k^2-\frac{a^{\prime\prime}}{a}\right)
v_{\bf k}=0,
$$
which is the equation of motion of a massless scalar field
in a quasi-de Sitter epoch. We can therefore make use of the results
present in Subsection 6.5 and Eq. (\ref{vv}) to  conclude that
on superhorizon scales the tensor modes scale like

$$
\left|v_{\bf k}\right|=\left(\frac{H}{2\pi}\right)\left(\frac{k}{aH}\right)
^{\frac{3}{2}-\nu_T},
$$
where 

$$
\nu_T\simeq \frac{3}{2}-\epsilon.
$$
Since fluctuations are (nearly)
frozen in on superhorizon scales,
a way of characterizing the tensor perturbations is to compute
the spectrum on scales larger than the horizon

\be   
{\cal P}_{T}(k)=\frac{k^3}{2\pi^2}\sum_{\lambda}\left|
h_{\bf k}
\right|^2=4\times 2\frac{k^3}{2\pi^2}\left|v_{\bf k}\right|^2.
\end{equation}
This gives the power spectrum on superhorizon scales

\begin{center}
\begin{tabular}{|p{13 cm}|}
\hline
%\\
$$
{\cal P}_{T}(k)=\frac{8}{\mpl^2}\left(\frac{H}{2\pi}\right)^2
\left(\frac{k}{aH}\right)^{n_T}\equiv A^2_{T}
\left(\frac{k}{aH}\right)^{n_T}
\label{ttt}
$$
\\
\hline
\end{tabular}
\end{center}
where we have defined the {\it spectral index} $n_{T}$ of the tensor 
perturbations 
as
\begin{center}
\begin{tabular}{|p{13 cm}|}
\hline
%\\   
$$
n_T=
\frac{d {\rm ln} \,{\cal P}_{T}}{d {\rm ln} \,k}=3-2\nu_T=
-2\epsilon.
$$
\\
\hline
\end{tabular}
\end{center}
The tensor perturbation is almost scale-invariant. Notice that
the amplitude of the tensor modes depends only on the value
of the Hubble rate during inflation. This amounts
to saying that it depends only on the energy scale $V^{1/4}$
associated to the inflaton potential. A detection of gravitational
waves from inflation will  therefore be a direct measurement
of the energy scale associated to inflation.

\subsection{The consistency relation}

The results obtained so far for the scalar and 
tensor perturbations allow one to predict a {\it consistency relation}
which holds for the models of inflation addressed in these
lectures, i.e., the models of inflation driven by
one-single field $\phi$. We define the tensor-to-scalar amplitude ratio to be

$$
r=\frac{\frac{1}{100} A_T^2}{\frac{4}{25}A_{\cal R}^2}=
\frac{\frac{1}{100} 8 \left(\frac{H}{2\, 
\pi\,\mpl}\right)^2}{\frac{4}{25}(2\epsilon)^{-1}\left(
\frac{H}{2\,\pi\,\mpl}\right)^2}=\epsilon.
$$
This means that 

\begin{center}
\begin{tabular}{|p{13 cm}|}
\hline
%\\
$$
r=-\frac{n_T}{2}
$$
\\
\hline
\end{tabular}
\end{center}
One-single models of inflation predict that during inflation
driven by a single scalar field, the ratio between the
amplitude of the tensor modes and that of the curvature perturbations
is equal to minus one-half of the tilt of the spectrum of tensor modes.
If this relation turns out to be
falsified by the future measurements of the CMB anisotropies, this
does not mean that inflation is wrong, but only that
inflation has not been driven by
only one field.

\subsection{From the inflationary seeds to the matter power spectrum}

As the curvature perturbations enter the causal horizon during radiation-
or matter-domination, they create density fluctuations $\delta\rho_{\bf k}$ 
via gravitational
attractions of the potential wells. The density contrast $\delta_{\bf k}=
\frac{\delta\rho_{\bf k}}{\overline\rho}$
can be deduced from the Poisson equation

$$
\frac{k^2\Phi_{\bf k}}{a^2}=-4\pi G\,\delta\rho_{\bf k}=
-4\pi G\,\frac{\delta\rho_{\bf k}}{\overline\rho}\,\overline\rho=
\frac{3}{2}\,H^2\,\frac{\delta\rho_{\bf k}}{\overline\rho}
$$
where $\overline\rho$ is the background average energy density.
This means that

$$
\delta_{\bf k}=\frac{2}{3}\,\left(\frac{k}{a H}\right)^2\,\Phi_{\bf k}.
$$
From this expression we can compute the power spectrum 
of matter density perturbations induced by inflation when they
re-enter the horizon during matter-domination:

$$
{\cal P}_{\delta\rho}=\langle\left|\delta_{\bf k}\right|^2\rangle=
A\,\left(\frac{k}{a H}\right)^n=\frac{2\pi^2}{k^3}\left(\frac{2}{5}\right)^2
 A^2_{\cal R}\left(\frac{k}{aH}\right)^4\,
\left(\frac{k}{aH}\right)^{n_{{\cal R}}-1}
$$
from which we deduce that matter perturbations scale linearly
with the wave-number and have a scalar tilt

$$
n=n_{{\cal R}}=1+2\eta-6\epsilon.
$$

The primordial spectrum ${\cal P}_{\delta\rho}$ is of course
reprocessed by gravitational
instabilities after the universe becomes matter-dominated. Indeed, 
as we have seen in Section 6, 
perturbations evolve after entering the horizon and the power spectrum will
not remain constant. To see how the density contrast is reprocessed
we have first to analyse how it evolves
on superhorizon scales before horizon-crossing. 
We use the following trick. Consider 
a flat universe with average energy density $\overline\rho$. The corresponding
Hubble rate is 

$$
H^2=\frac{8\pi G}{3}\,\overline\rho.
$$

A small
positive fluctuation $\delta\rho$ will cause the universe to be
closed:

$$
H^2=\frac{8\pi G}{3}\left(\overline\rho+\delta\rho\right)-
\frac{k}{a^2}.
$$
Substracting the two equations we find

$$
\frac{\delta\rho}{\rho}=\frac{3}{8\pi G}\frac{k}{a^2\rho}\sim
\left\{\begin{array}{cc}
a^2 &{\rm RD}\\
a & {\rm MD}\end{array}\right.
$$
Notice that $\Phi_{\bf k}\sim \delta\rho a^2/k^2\sim(\delta\rho/\rho)\rho
a^2/k^2=$ constant for both RD and MD which confirms our previous findings.

When the matter densities enter the horizon, they 
do not increase appreciably before 
matter-domination because before matter-domination pressure
is too large and does not allow the matter inhomogeneities to grow.
On the other hand, the suppression of growth due to radiation
is restricted to scales smaller than the horizon, while large-scale
perturbations remain unaffected. This is why
the horizon size at equality sets an important scale for structure growth:

$$
k_{\rm EQ}=H^{-1}\left(a_{\rm EQ}\right)\simeq
0.08\,h\,{\rm Mpc}^{-1}.
$$
Therefore, perturbations with $k\gg k_{\rm EQ}$ are perturbations
which have entered the horizon before matter-domination and have remained
nearly constant till equality. This means that they are suppressed
with respect to those perturbations having $k\ll k_{\rm EQ}$ by a factor
$(a_{\rm ENT}/a_{\rm EQ})^2=(k_{\rm EQ}/k)^2$. 
If we define the 
 transfer function $T(k)$ by the relation
${\cal R}_{\rm final}=T(k)\,{\cal R}_{\rm initial}$ we find therefore
that roughly speaking
$$
T(k)=\left\{\begin{array}{cc}
1 &k\ll k_{\rm EQ},\\
(k_{\rm EQ}/k)^2 &k\gg k_{\rm EQ}. \end{array}\right.
$$
The corresponding power spectrum will be

$$
{\cal P}_{\delta\rho}(k)\sim\left\{\begin{array}{cc}
\left(\frac{k}{aH}\right) & k\ll k_{\rm EQ},\\
 \left(\frac{k}{aH}\right)^{-3} & k\gg k_{\rm EQ}. \end{array}\right.
$$
Of course, a more careful computation needs to include many other effects
such as neutrino free-streaming, photon diffusion and the diffusion of baryons
along with photons. It is encouraging, however, that this rough estimate
turns out to be confirmed by present data on large-scale structures \cite{llbook}.

The next step would be to investigate how the primordial perturbations generated by inflation flow into the CMB
to produce their anisotropies.

\section{From inflation to large-angle CMB anisotropy}

Temperature fluctuations in the CMB arise due to various distinct
physical effects: first of all due to our peculiar velocity with respect to
the cosmic rest frame; fluctuations in the gravitational potential
on the last scattering surface appear also at the last scattering period; fluctuations are also intrinsic to the radiation
field itself on the last scattering surface.  Finally, there is the contribution from
the evolution of the anisotropies from the last scattering surface till today
(which we shall neglect from now on).

The second effect, known as the Sachs--Wolfe effect is the dominat contribution
to the anisotropy on large-angular scales, $\theta\gg\theta_{\rm HOR}\sim
 1^\circ$. The last three
effects provide the dominant contributions to the anisotropy on small-angular 
scales,
 $\theta\ll 1^\circ$.

\subsection{Sachs--Wolfe plateau}

We consider first the temperature fluctuations on large-angular scales
that arise due to the Sachs--Wolfe effect.  They provide
a probe of the original spectrum of primeval fluctuations produced during 
inflation. One can show that

\begin{equation}
\label{sw1}
\left(\frac{\delta T}{T}\right)_{*}=\int_0^{x_{\rm LS}}\,\frac{t}{a}
\frac{d^2\Phi}{dx^2}\,dx.
\ee
The photon trajectory is $a d{\bf x}/dt={\bf n}$. Using $a\sim t^{2/3}$ gives

$$
x(t)=\int_t^{t_0}\frac{dt^\prime}{a}=3\left(\frac{a_0}{t_0}-
\frac{t}{a}\right).
$$
Integrating by parts Eq. (\ref{sw1}), we finally find

$$
\left(\frac{\delta T}{T}\right)_{*}=\frac{1}{3}\left[
\Phi({\bf x}_{\rm LS})-
\Phi(0)\right]+{\bf e}\cdot\left[{\bf v}(0,t_0)-{\bf v}(
{\bf x}_{\rm LS},t_{\rm LS})\right].
$$
The potential at our position contributes only to
the unobservable monopole and can be dropped. On scales outside the horizon,
${\bf v}=-t\nabla\Phi\sim 0$. The remaining term is the
Sachs--Wolfe effect

$$
\frac{\delta T({\bf e})}{T}=\frac{1}{3}\Phi({\bf x}_{\rm LS})=
\frac{1}{5}{\cal R}({\bf x}_{\rm LS}).
$$
This relation has been obtained as follows. The co-moving curvature 
perturbation is given during the radiation phase by ${\cal R}=\psi+H\delta\rho/ \dot{\rho}=\psi-1/3\delta\rho_\gamma/\rho_\gamma$. Einstein equations set
$\psi=\Phi$ and $ \delta\rho_\gamma/\rho_\gamma=-2\Phi$ on super-horizon scales. Therefore ${\cal R}=5/3\Phi$ beyond the horizon. 

At large angular scales, the theory of cosmological perturbations
predicts a remarkably simple formula relating the CMB anisotropy to
the curvature perturbation generated during inflation.

We have seen previously that 
the 
temperature anisotropy is commonly expanded  in spherical harmonics
$
\frac{\Delta T}{T}(x_0,\tau_0,{\bf n})=\sum_{\ell m}
a_{\ell,m}(x_0)Y_{\ell m}({\bf n}),
$
where $x_0$ and $\tau_0$ are our position and the preset time, respectively,
 ${\bf n}$ is the
direction of observation, $\ell'$s are the
different multipoles,  and
$
\langle a_{\ell m}a^*_{\ell'm'}\rangle=\delta_{\ell,\ell'}\delta_{m,m'} C_\ell
$,
where the deltas are due to the fact that the process that created
the anisotropy is statistically isotropic. 
The $C_\ell$'s are the so-called CMB power spectrum.
For homogeneity and isotropy, the $C_\ell$'s are neither a function
of $x_0$, nor of $m$.
The two-point-correlation function is related to the $C_l$'s
according to Eq. (\ref{j}).

For adiabatic perturbations we have seen that on large scales,
larger than the horizon on the last scatteringsurface (corresponding
to angles larger than $\theta_{\rm HOR}\sim 1^\circ$)
$\delta T/T=
\frac{1}{3}\Phi({\bf x}_{\rm LS})$.
In Fourier transform
\begin{equation}
\frac{\delta T({\bf k},\tau_0,{\bf n})}{T}=
\frac{1}{3}\Phi_{{\bf k}}\,e^{i\,{\bf k}\cdot{\bf n}(\tau_0-
\tau_{{\rm LS}})}.
\end{equation}
Using the decomposition
\begin{equation}
\exp(i\,{\bf k}\cdot{\bf n}(\tau_0-\tau_{{\rm LS}}))=
\sum_{\ell=0}^\infty (2\ell+1) i^\ell
j_\ell(k(\tau_0-\tau_{{\rm LS}})) P_{\ell}({\bf k}\cdot{\bf n})
\end{equation}
where $j_\ell$ is the spherical Bessel function of order $\ell$ and 
substituting, we get
\begin{eqnarray}
&&\Big<\frac{\delta T(x_0,\tau_0,{\bf n})}{T}\frac{\delta
T(x_0,\tau_0,{\bf n'})}{T}\Big>=\\ \nonumber &&=\frac{1}{V}\int
d^3x \Big<\frac{\delta T(x_0,\tau_0,{\bf n})}{T}\frac{\delta
T(x_0,\tau_0,{\bf n}')}{T}\Big>=\\ \nonumber
&&=\frac{1}{(2\pi)^3}\int d^3k \Big<\frac{\delta
T({\bf k},\tau_0,{\bf n})}{T}\left(\frac{\delta
T({\bf k},\tau_0,{\bf n}')}{T}\right)^*\Big>=\\ \nonumber
&&=\frac{1}{(2\pi)^3}\int d^3k
\Big(\Big<\frac{1}{3}|\Phi|^2\Big>
\sum_{\ell,\ell'=0}^{\infty}
(2\ell+1)(2\ell'+1)j_\ell(k(\tau_0-\tau_{\rm LS}))\nonumber\\
&&j_{\ell'}(k(\tau_0-\tau_{{\rm LS}}))
P_\ell({\bf k}\cdot{\bf n}) P_{\ell'}({\bf k}'\cdot{\bf n}')\Big)
\end{eqnarray}
Inserting $P_\ell({\bf k}\cdot{\bf n})=\frac{4\pi}{2\ell+1}\sum_m
Y^*_{lm}({\bf k})Y_{\ell m}({\bf n})$ and analogously for
$P_\ell({\bf k}'\cdot{\bf n}')$, integrating over the directions
$d\Omega_k$ generates $\delta_{\ell\ell'}\delta_{mm'}\sum_m
Y^*_{\ell m}({\bf n})Y_{\ell m}({\bf n}')$. Using as well $\sum_m
Y^*_{\ell m}({\bf n})Y_{\ell m}({\bf n}')=\frac{2\ell+1}{4\pi}
P_\ell({\bf n}\cdot {\bf n}')$, we
get
\begin{eqnarray}
&&\Big<\frac{\delta T(x_0,\tau_0,{\bf n})}{T}\frac{\delta
T(x_0,\tau_0,{\bf n}')}{T}\Big>\\ \nonumber &&=\Sigma_\ell
\frac{2\ell+1}{4\pi}P_\ell({\bf n}\cdot{\bf n}') 
\frac{2}{\pi}\int \frac{dk}{k}
\Big<\frac{1}{9}|\Phi|^2\Big> k^3 j^2_\ell(k(\tau_0-\tau_{\rm LS})).
\end{eqnarray}
Comparing this expression with that for the $C_\ell$, we get the
expression for the $C^{{\rm AD}}_\ell$, where the suffix ``AD'' stands for
adiabatic: 
\begin{equation}
C^{\rm AD}_\ell=\frac{2}{\pi}\int \frac{dk}{k} \Big<\frac{1}{9}\left|\Phi
\right|^2\Big> k^3
j^2_\ell(k(\tau_0-\tau_{\rm LS}))
\end{equation}
which is valid for $2\leq \ell\ll
(\tau_0-\tau_{\rm LS})/\tau_{\rm LS}\sim 100$.

If we generically indicate by 
$\langle|\Phi_{\bf k}|^2\rangle k^3=A^2\,(k\tau_0)^{n-1}$, 
we can perform the integration and
get
\begin{equation}
\frac{\ell(\ell+1)C^{\rm AD}_\ell}{2\pi}
=\left[
\frac{\sqrt{\pi}}{2}\ell(\ell+1)\frac{\Gamma(\frac{3-n}{2})
\Gamma(\ell+\frac{n-)}{2})}{\Gamma\left(\frac{4-n}{2}\right)\Gamma
\left(\ell+
\frac{5-n}{2}\right)}\right]
\frac{A^2}{9}\left(\frac{H_0}{2}\right)^{n-1}.
\end{equation}
For $n\simeq 1$ and $\ell\gg 1$, we can approximate this expression to

\begin{equation}
\frac{\ell(\ell+1)C^{\rm AD}_l}{2\pi}=\frac{A^2}{9}.
\label{Clad}
\end{equation}
This result shows that inflation predicts
a very flat spectrum  for  low $\ell$.
Furthermore, 
since inflation predicts $\Phi_{\bf k}=\frac{3}{5}{\cal R}_{\bf k}$, we find
that 

\begin{equation}
\pi\,\ell(\ell+1)C^{\rm AD}_l=\frac{A_{\cal R}^2}{25}=
\frac{1}{25}\frac{1}{2\,\mpl^2\,\epsilon}\left(\frac{H}{2\pi}\right)^2.
\end{equation}

WMAP5 data imply that $\frac{\ell(\ell+1)C^{\rm AD}_l}{2\pi}\simeq 10^{-10}$
or
\begin{center}
\begin{tabular}{|p{13 cm}|}
\hline
%\\
$$
\left(\frac{V}{\epsilon}\right)^{1/4}\simeq 6.7\times 10^{16}\,{\rm GeV}
$$
\\
\hline
\end{tabular}
\end{center}

\subsection{Acoustic peaks}
This part is heavily taken from \cite{zal}. We thank M. Zaldarriaga for granting permission.
To be able to calculate the power spectrum of the anisotropies even on 
angular scales
larger than $1^\circ$, we need
to consider the evolution of the photon anistoropies. 
As we already mentioned, before recombination Thomson
scattering was very efficient. As a result it is a good approximation
to treat photons and baryons as a single fluid. 
This treatment is called the tight-coupling approximation and will allow
us to evolve the perturbations until recombination.

The equation for the photon density perturbations for one Fourier
 mode of wave-number $k$ is that of a forced and damped
harmonic oscillator
\begin{eqnarray}\label{fdosc}
 && \ddot{\delta}_\gamma+{\dot{R} \over (1+R)} \dot{\delta}_{\gamma}
  +k^2 c^2_{s} \delta_{\gamma}=F, \nonumber \\
&&F=4[\ddot \psi +{\dot R \over (1+R)} \dot \psi - {1 \over 3} k^2
  \Phi], \nonumber \\
&&\dot{\delta}_\gamma= -{4 \over 3 } k v_\gamma + 4 \dot \psi.
\end{eqnarray}
The photon--baryon fluid can sustain acoustic oscillations.
The inertia is provided by the baryons, while the pressure is provided
by the photons. The sound speed is $c_s^2=1/3(1+R)$, with
$R=3\rho_b/4\rho_{\gamma}=31.5\ (\Omega_b h^2)
(T/2.7)^{-4}[(1+z)/10^3]^{-1}$.  As the baryon
fraction goes down, the sound speed approaches $c_s^2\rightarrow 1/3$. 
The third equation above is the continuity equation. 

As a toy problem, we shall solve Eq. (\ref{fdosc}) under 
some simplifying assumptions. If we
consider a matter-dominated universe, the driving force becomes a
constant, $F=-4/3 k^2 \Phi$, 
because the gravitational potential remains constant in time. 
We neglect anisotropic stresses so that $\psi=\Phi$, and,  
furthermore, we neglect the time dependence of $R$.  
Equation (\ref{fdosc}) becomes
that of a harmonic oscillator that can be trivially solved. This is
a very simplified picture, but it captures most
of the relevant physics we want to discuss. 

To obtain the final solution we need again to specify the initial
conditions. we shall restrict ourselves to adiabatic initial
conditions, the most natural outcome of inflation. In our context this
means that initially $\Phi=\psi=\Phi_0$, $\delta_\gamma=-8/3 \Phi_0$, 
and $v_\gamma=0$. We have denoted $\Phi_0$ the initial amplitude of
the potential fluctuations. We shall take $\Phi_0$ to be a Gaussian 
random variable with power spectrum $P_{\Phi_0}$. 

We have made enough approximations that the evaluation of the sources
in the integral solution has become trivial. The solution for
the density and velocity of the photon fluid at recombination is
\begin{eqnarray}\label{solution1}
\left({\delta_{\gamma} \over 4}
+\Phi\right) |_{\rm LS}&=&{\Phi_0 \over 3}(1+3R)\cos (k c_s \tau_{\rm LS})-\Phi_0 
R, \nonumber \\
v_\gamma |_{\tau_{\rm LS}}&=& -\Phi_0 (1+3R) c_s \sin(k c_s \tau_{\rm LS}).
\end{eqnarray}
Equation (\ref{solution1}) is the solution for a single Fourier
mode. All quantities have an additional spatial dependence
($e^{i\bf k \cdot \bf x}$), which we have not included in order to make the
notation more compact. 
With that additional term the solution we have is
\begin{eqnarray}
  \frac{\delta T}{T}({\bf n})&=&e^{ikD_{\rm LS}\cos\theta} S \nonumber \\
S&=&\Phi_0 {(1+3R) \over 3}
[\cos (k c_s \tau_{\rm LS})- {3 R \over (1+3R)}, \nonumber \\
&& - i \sqrt{3 \over 1+R} 
\cos \theta \sin(k c_s
  \tau_{\rm LS})],
\label{integapprox1mode}
\end{eqnarray}
where we have neglected the $\Phi$ on the left-hand side because it is
a constant.  
We have introduced $\cos \theta$, the cosine of the angle between
the direction of observation and the wavevector ${\bf k}$; for example, 
${\bf k} \cdot {\bf  x}=k D_{\rm LS} \cos \theta$ . The term
proportional to $\cos \theta$ is the Doppler contribution. 

Once the temperature perturbation produced by one Fourier mode has
been calculated, we need to expand it into spherical harmonics.
The power spectrum of temperature
anisotropies is expressed in terms of the $a_{lm}$ coefficients as
$C_{T\ell}=\sum_m |a_{\ell m}|^2$. The contribution to $C_{Tl}$ from each
Fourier mode is weighted by the amplitude of primordial fluctuations
in this mode, characterized by the power spectrum of $\Phi_0=3/5{\cal R}$, 
$P_{\Phi_0}=A k^{-3}$ as dictated by inflation. 
In practice, fluctuations on angular scale
$\ell$ receive most of their contributions from wavevectors around
$k_*=\ell/D_{\rm LS}$, so roughly the amplitude of the power spectrum 
at multipole $\ell$ is given by the value of the sources in Eq.
(\ref{solution1}) at $k_*$.

After summing the contributions from all modes, the power spectrum is
roughly given by

\begin{eqnarray}\label{approxcl}
  \ell(\ell+1)C_{Tl}&\approx&A \{[{(1+3R)\over 3} \cos(k_*c_s\tau_{\rm LS})-R]^2+ 
\nonumber \\
&& {(1+3R)^2
  \over 3} c_s^2  
  \sin^2(k_*c_s\tau_{\rm LS})\}.
\end{eqnarray}
Equation (\ref{approxcl}) can be used to understand the basic
features in the CMB power spectra. 
The baryon drag on the photon--baryon fluid reduces its sound speed below 
$1/3$ and makes the monopole
contribution dominant (the one proportional to $\cos(k_*c_s\tau_{\rm LS}$). 
Thus, the $C_{Tl}$ spectrum peaks where the
monopole term peaks, $k_*c_s\tau_{\rm LS}=\pi,2\pi,3\pi,\cdots$, which
correspond to $\ell_{\rm peak}=n\pi D_{\rm LS}/c_S\tau_{\rm LS}$. 

It is very important to understand the origin of the acoustic
peaks. In this model the universe is
filled with standing waves; all modes of wave-number $k$ are in
phase, which leads to the oscillatory terms. The sine and cosine in Eq.
(\ref{approxcl}) originate in the time dependence of the modes. 
Each mode $\ell$
receives contributions preferentially from Fourier modes of a
particular wavelength $k_*$ (but pointing in all directions), 
so to obtain peaks in $C_\ell$,  
it is crucial that all modes of a given $k$ be in phase.
If this is not the case, the features in the $C_{T\ell}$ spectra will be
blurred and can even disappear. This is what happens when one considers the
spectra produced by topological defects.
The phase coherence of all
modes of a given wave-number 
can be traced to the fact that perturbations were produced very
early on and had wavelengths larger than the horizon during many expansion
times. 

There are additional physical effects we have neglected. 
 The universe was radiation dominated early on,
and modes of wavelength smaller and bigger than the horizon at
matter-radiation equality behave differently. During the
radiation era the perturbations in the photon--baryon
fluid are the main source for the gravitational potentials which decay once a mode enters into the horizon. The
gravitational potential decay acts as a driving force for the oscillator
in Eq. (\ref{fdosc}), so a feedback loop is established. As a
result, the acoustic oscillations for 
modes that entered the horizon before matter-radiation
equality have a higher amplitude. In the $C_{T\ell}$ spectrum the
separation between modes that experience this feedback and those
that do not occurs at $\ell\sim D_{\rm LS}/ \tau_{\rm LS}$. Larger $\ell$ values
receive their contributions from modes that entered the horizon
before matter-radiation equality. Finally, when a mode is inside the
horizon during the radiation era the gravitational potentials decay. 

There is a competing effect, Silk damping,  that reduces the 
amplitude of the large-$l$ modes. The photon--baryon fluid
is not a perfect fluid. Photons have a finite mean free path and
thus can random-walk away from the peaks and valleys 
of the standing waves.
Thus perturbations of wavelength comparable to or smaller than the
distance the photons can random-walk get damped. This
effect can be modelled by multiplying Eq. \ref{integapprox1mode}
by $\exp(-k^2/k_s^2)$, with $k^{-1}_{s}\propto \tau_{\rm LS}^{1/2} (\Omega_b
h^2)^{-1/2}$. Silk damping is important for multipoles of order 
$\ell_{\rm Silk}\sim
k_s D_{\rm LS}$.
Finally, the last scatteringsurface has a finite width. 
Perturbations
with wavelength comparable to this width get smeared out due to
cancellations along the line of sight. This
effect introduces an additional damping with a characteristic scale
$k^{-1}_w \propto \delta\tau_{\rm LS}$.

The location of the first peak is by itself a measurement of the
geometry of the universe. In fact, photons propagating on geodesics
from the last scattering surface to us feel the spatial geometry, 
whose properties we learned are dictated by $\Omega_0$. In fact, the location
of the first peak is given by $\ell_1\simeq 220/\sqrt{\Omega_0}$. WMAP5
gives $\Omega_0=1.00^{+0.07}_{-0.03}$. This tells us that the spatial
(local) geometry of the universe is flat. This is precisely what
inflation predicts. 

\subsection{The polarization of the CMB anisotropies}

The anisotropy field is characterized by a $2\times 2$ intensity 
tensor $I_{ij}$. For convenience, we normalize this tensor so that it represents 
the fluctuations in
units of the mean intensity ($I_{ij}=\delta I /I_0$). 
The intensity tensor is a
function of direction on the sky, ${\bf n}$, and  two directions
perpendicular to ${\bf n}$ that are  used to define its components
(${\bf  e}_1$,${\bf  e}_2$).
The Stokes parameters $Q$ and $U$ are defined as
$Q=(I_{11}-I_{22})/4$ and $U=I_{12}/2$, while the temperature 
anisotropy is
given by $T=(I_{11}+I_{22})/4$ (the factor of $4$ relates fluctuations
in the intensity with those in the temperature, $I\propto T^4$).  
When representing polarization using ``rods'' in a map, 
the magnitude is given by $P=\sqrt{Q^2+U^2}$, and the 
orientation makes 
an angle $\alpha={1\over 2}\arctan({U/Q})$ with ${\bf  e}_1$.
In principle the fourth  
Stokes parameter $V$ that describes circular polarization is needed, 
but we ignore it because it cannot
be generated through Thomson scattering, so the CMB is not expected to
be circularly polarized.   
While the temperature is invariant
under a right-handed rotation in the plane perpendicular to direction
${\bf n}$,
$Q$ and $U$ transform under rotation by an angle $\psi$ as
\begin{equation}
(Q\pm iU)'({\bf n})=e^{\mp 2i\psi}(Q\pm iU)({\bf n}),
\end{equation}
where ${\bf  e}_1^{\prime}=\cos \psi\ {\bf  e}_1+\sin\psi\ 
{\bf  e}_2$ 
and ${\bf  e}_2^{\prime}=-\sin \psi\ {\bf  e}_1+\cos\psi\ 
{\bf  e}_2$. The quantities $Q\pm iU$ are said to be spin 2. 

We already mentioned  that the statistical properties of the
radiation field are usually described in terms of the spherical harmonic
decomposition of the maps. This basis, basically the Fourier basis, is
very natural because the statistical properties of 
anisotropies are rotationally invariant.  
The standard spherical harmonics are not the
appropriate basis for $Q\pm iU$ because they are spin-2 variables, but 
generalizations (called $_{\pm 2} Y_{lm}$) exist. We can expand 
\begin{eqnarray}
(Q\pm iU)({\bf n})&=&\sum_{\ell m} 
a_{\pm 2,\ell m}\;_{\pm 2}Y_{\ell m}({\bf n}). 
\label{Pexpansion}
\end{eqnarray}
Here $Q$ and $U$ are defined at each direction $\hat {{\bf n}}$
with respect to the spherical coordinate system $({\bf e}_\theta,
{\bf e}_\phi)$.
To ensure that $Q$ and $U$ are real, 
the expansion coefficients 
must satisfy $a_{-2,\ell m}^*=a_{2,\ell-m}$. The equivalent relation 
for the temperature coefficients is $a_{T,\ell m}^*=a_{T,\ell-m}$.
Instead of $a_{\pm 2,\ell m}$, it is convenient to introduce their
linear combinations 
$a_{E,\ell m}=-(a_{2,\ell m}+a_{-2,\ell m})/2$ and
$a_{B,\ell m}=i(a_{2,\ell m}-a_{-2,\ell m})/2$. 
We define two quantities in real space,
$E({\bf n})=\sum_{\ell,m}a_{E,\ell m}  Y_{\ell m}({\bf n})$ and 
$B({\bf n})=\sum_{\ell,m}a_{B,\ell m}  Y_{\ell m}({\bf n})$. Here $E$ and
$B$ completely specify the linear polarization field. 

The temperature is a
scalar quantity
under a rotation of the coordinate system,
$T^{\prime}({\bf n}^{\prime}={\bf \cal R} {\bf n})
=T({\bf n})$, where  $\bf {\cal
R}$ is the rotation matrix. We denote with a
prime the quantities in the transformed coordinate system. While $Q\pm
i U$ are spin 2,
$E({\bf n})$ and $B({\bf n})$ are 
invariant under rotations. Under parity, however, $E$ and $B$ behave
differently, $E$ remains unchanged, while $B$ changes sign.

%%%%%%%%%%%%%%%%%%%%%%%%%%%%%%%%%%%%%%%%%%%%%%%%%%%%%%%%%%%%%%%%
\begin{figure}
\centering\includegraphics[width=.5\linewidth]{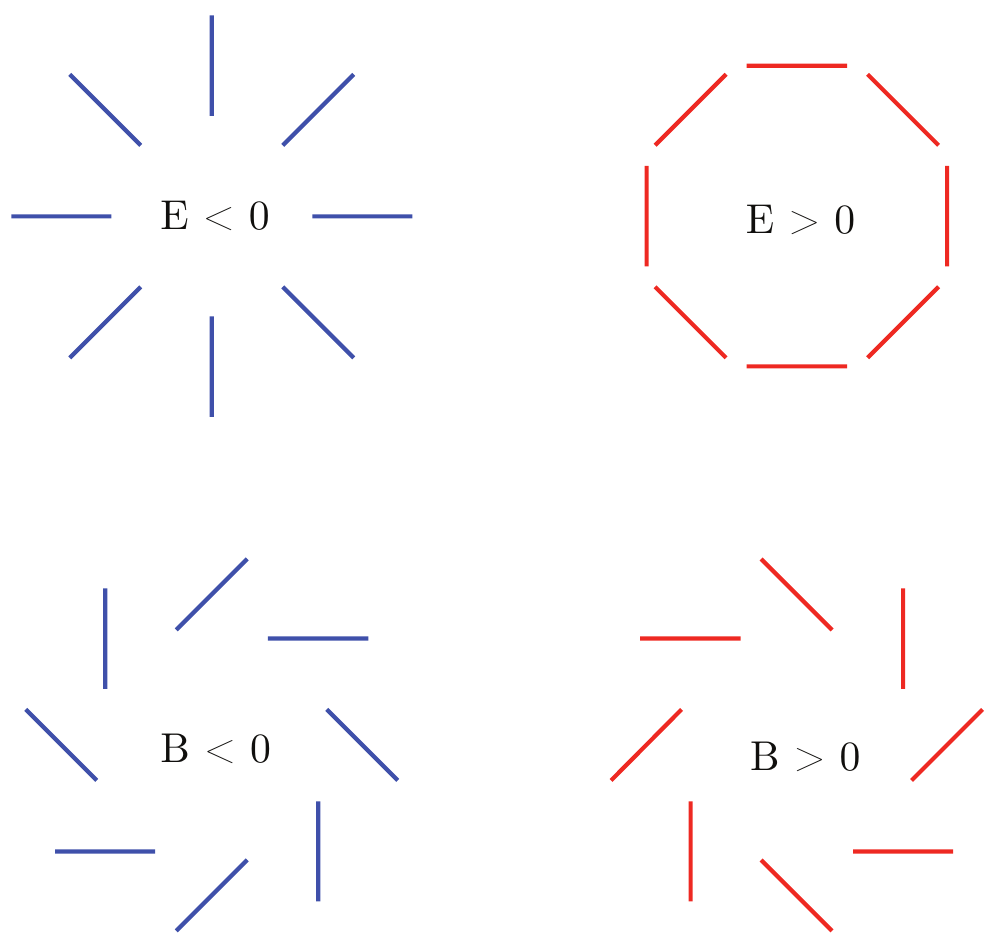}
\caption{Examples of $E$- and $B$-mode patterns of 
polarization}
\end{figure}
%%%%%%%%%%%%%%%%%%%%%%%%%%%%%%%%%%%%%%%%%%%%%%%%%%%%%%%%%%%%%%%%

To characterize the statistics of the CMB perturbations, 
only four power spectra are needed, 
those for $T$, $E$, $B$ and the cross correlation between $T$ and $E$.
The cross correlation between $B$ and $E$ or $B$ and 
$T$ vanishes if there are no parity-violating interactions because 
$B$ has the opposite parity to $T$ or $E$. 
The power spectra are defined as the rotationally invariant quantities
$C_{T\ell}={1\over 2\ell+1}\sum_m \langle a_{T,\ell m}^{*} a_{T,\ell m}\rangle$, 
$C_{E\ell}={1\over 2\ell+1}\sum_m \langle a_{E,\ell m}^{*} a_{E,\ell m}\rangle$, 
$C_{B\ell}={1\over 2\ell+1}\sum_m \langle a_{B,\ell m}^{*} a_{B,\ell 
m}\rangle$, 
and
$C_{C\ell}={1\over 2\ell+1}\sum_m \langle a_{T,\ell m}^{*}a_{E,\ell m}\rangle$.
The brackets $\langle \cdots \rangle$ denote ensemble averages.

Polarization is generated by Thomson scattering between photons and
electrons, which means that polarization cannot be generated after
recombination (except for re-ionization, which we shall discuss later). 
But Thomson scattering is
not enough. The radiation incident on the electrons must also be
anisotropic. In fact, its intensity needs to have a quadrupole moment. This
requirement of having both Thomson scattering and anisotropies is what
makes polarization relatively small. After recombination, anisotropies
grow by free streaming, but there is no scattering to generate
polarization. Before recombination there were so many scatterings
that they erased any anisotropy present in the photon--baryon fluid.

%%%%%%%%%%%%%%%%%%%%%%%%%%%%%%%%%%%%%%%%%%%%%%%%%%%%%%%%%%%%%%%%
\begin{figure}
\centering\includegraphics[width=.5\linewidth]{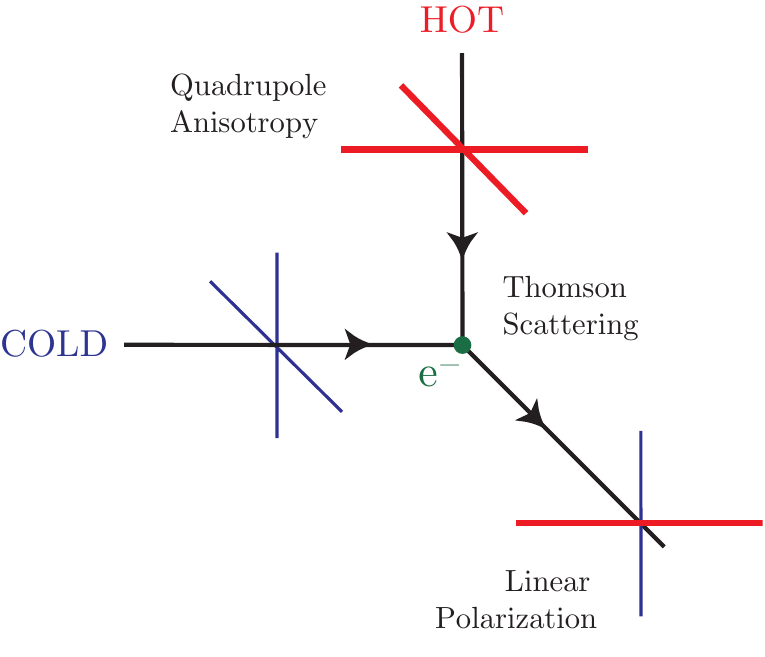}
\caption{Thomson scattering of radiation where quadrupole anisotropy generates linear
polarization}
\end{figure}
%%%%%%%%%%%%%%%%%%%%%%%%%%%%%%%%%%%%%%%%%%%%%%%%%%%%%%%%%%%%%%%%

In the context of anisotropies induced by density perturbations,
velocity gradients in the photon--baryon fluid are responsible for the
quadrupole that generates polarization.  Let us consider a scattering
occurring at position $\bi x_0$: the scattered photons came from a
distance of order the mean free path ($\lambda_T$) away from this
point. If we are considering photons traveling in direction $\bi
{\hat n}$, they roughly come from $\bi x=\bi x_0 + \lambda_T \bi {\hat
  n}$. The photon--baryon fluid at that point was moving at velocity
$\bi v(\bi x)\approx \bi v(\bi x_0)+\lambda_T {\bi {\hat n}}_i
\partial_i {\bi v}(\bi x_0)$. Due to the Doppler effect the
temperature seen by the scatterer at $\bi x_0$ is $\delta T(\bi
x_0,\bi {\hat n})= \bi {\hat n}\cdot [\bi {v}(\bi x)-\bi v(\bi
x_0)]\approx\lambda_T {\bi {\hat n}}_i{\bi {\hat n}}_j \partial_i {\bi
  v}_j(\bi x_0)$, which is quadratic in $\bi{\hat n}$ (i.e., it has a
  quadrupole). Velocity gradients in the photon--baryon fluid
lead to a quadrupole component of the intensity distribution, which,
through Thomson scattering, is converted into polarization.  

The polarization of the scattered radiation field, expressed in terms
of the Stokes parameters $Q$ and $U$, 
is given by $(Q+iU) \propto \sigma _T\int d\Omega
^{\prime} ({\bf m}\cdot \hat {{\bf n}}^{\prime })^2 T(\hat {{\bf n}%
}^{\prime })$ $ \propto \lambda _p{\bf m}^i{\bf m}^j\partial
_iv_j|_{\tau_{\rm LS}}$, 
where $\sigma _T$ is the Thomson scattering cross-section and we have
written the
scattering matrix as $P({\bf m},\hat {{\bf n}}^{\prime })=-3/4\sigma _T({\bf %
m}\cdot \hat {{\bf n}}^{\prime })^2$, with ${\bf m}=\hat {{\bf e}}_1+i\hat {%
{\bf e}}_2$ . In the last step, we integrated over all directions of the
incident photons $\hat {{\bf n}}^{\prime }$.
As photons decouple from the baryons, their mean free path grows very
rapidly, so a more careful analysis is needed to obtain the final
polarization: 
\begin{eqnarray}
(Q+iU)(\hat {{\bf n}})\approx \epsilon \delta \tau _{\rm LS}{\bf m}^i
{ \bf m}^j\partial
_iv_j|_{\tau _{\rm LS}}, 
\label{polapprox}
\end{eqnarray}
where $\delta \tau _{\rm LS}$ is the width of the last scattering surface and
gives a measure of the distance that photons travel between their last two
scatterings, and $\epsilon$ is a numerical constant that depends on the
shape of the visibility function. 
The appearance of ${\bf m}^i {\bf m}^j$ in Eq. (\ref{polapprox}) ensures
that $(Q+iU)$ transforms correctly under rotations of $(\hat {{\bf e}%
}_1,\hat {{\bf e}}_2)$. 

If we evaluate Eq. 
(\ref{polapprox}) for each Fourier mode and combine them
to obtain the total power, we get the equivalent of Eq.
(\ref{approxcl}),
\begin{eqnarray}\label{approxcel}
  \ell(\ell+1)C_{E\ell}\approx A \epsilon^2 (1+3R)^2 (k_*\delta\tau_{\rm LS})^2  
\sin^2(k_*c_s\tau_{\rm LS}),
\end{eqnarray}
where we are assuming $n=1$ and that $\ell$ is large enough that factors
like $(\ell+2)! /(\ell-2)!\approx \ell^4$. The extra $k_*$ in Eq.
(\ref{approxcel}) originates in the gradient in Eq.
(\ref{polapprox}). 
The large-angular scale
polarization is greatly suppressed by the $k\delta\tau_{\rm LS}$ factor. 
Correlations over large angles can only be created by the long-wavelength 
perturbations, but these cannot produce a large
polarization signal because of the tight coupling between photons and
electrons prior to recombination. Multiple scatterings make the
plasma very homogeneous; only wavelengths that are small enough to
produce anisotropies over the  mean free path of the photons will give
rise to a significant quadrupole in the temperature distribution, and
thus to polarization. Wavelengths much smaller than the mean free path
decay due to photon diffusion (Silk damping) and so are unable to
create a large quadrupole and polarization. As a result polarization
peaks at the scale of the mean free path. 

On sub-degree  angular 
scales, temperature, polarization, and the cross-correlation power
spectra show
acoustic oscillations.
In the polarization and cross-correlation spectra the peaks are much sharper. 
The polarization is produced
by  velocity gradients of the photon---baryon fluid
at the last scatteringsurface. 
The temperature receives
contributions from density and velocity perturbations, 
and  the oscillations in each partially
cancel one another, making the features in the temperature spectrum less
sharp. The dominant contribution to the temperature comes from the 
oscillations in the density [Eq. (\ref{solution1})], 
which are out of phase with the velocity.
This explains the difference in location between the temperature and
polarization peaks. The extra gradient in the polarization signal,
Eq. 
(\ref{polapprox}), explains why its overall amplitude peaks at a smaller
angular scale. 

Now, as photons travel in the metric perturbed by a GW 
[$ds^2=a^2(\tau)$ $[-d\tau^2 $ $ +(\delta_{ij}+ h^{T}_{ij})dx^i dx^j]$],
they get redshifted or blueshifted depending on their direction of
propagation relative to the direction of propagation of the GW
and the polarization of the GW. For example, for a
GW travelling along the $z$ axis, the frequency shift is
given by 

\begin{equation}
{1 \over \nu}{d \nu \over d\tau}= {1 \over 2}\ \hat
{n}^i\hat {n}^j {\dot h}^{T(\pm)}_{ij}= {1 \over 2}\ (1-\cos^2
\theta)e^{\pm i 2\phi}\ \ \dot h_t\ \exp(i\bf k\cdot \bf x),
\end{equation}
 where
$(\theta,\phi)$ describe the direction of propagation of the photon,
the $\pm$ correspond to the different polarizations of the GW, 
and $h_t$ gives the time-dependent amplitude of the GW.
During the matter-dominated era, for example, 
$h_t= 3 j_1(k\tau)/k\tau$: time changes in the metric
lead to frequency shifts (or equivalently shifts in the temperature of
the black body spectrum). Notice that the angular dependence of this
frequency shift is quadrupolar in nature. As a result, the temperature
fluctuations induced by this effect as photons travel between
successive scatterings before recombination 
produce a quadrupole intensity distribution, which,
through Thomson scattering, lead to polarization. 
Both $E$ and $B$ 
power spectra are generated by GW. 
The current push to improve polarization measurements
follows from the fact that density perturbations, to linear order in
perturbation theory, cannot create any $B$-type polarization. 
As a rough rule of thumb, the amplitude of the peak in the $B$-mode
power spectrum for GW is

\begin{center}
\begin{tabular}{|p{13 cm}|}
\hline
%\\
$$
[{\ell(\ell+1)C_{Bl}/ 2\pi }]^{1/2}=0.024 ({V^{1/4}/ 10^{16}{\rm
GeV}})^2 \mu \rm K
 $$
\\
\hline
\end{tabular}
\end{center}
where 

\begin{equation}
V^{1/4}\simeq 6.7\,r^{1/4}\,\times 10^{16}\,{\rm GeV}
\end{equation}
is the energy scale of inflation. A future experiment like CMBPol \cite{CMBPol} can probe values of $r$ as small as
$10^{-2}$, corresponding to an inflation energy scale of about $2\times \times 10^{16}$ GeV. Furthermore, using the
consistency relation $r=\epsilon$ valid in one-single field models of inflation, one deduces that 

\begin{equation}
\frac{\Delta \phi}{m_{\rm Pl}}\simeq \left(\frac{r}{10^{-2}}\right)^{1/2},
\end{equation}
meaning that a future measurement of the $B$-mode of CMB polarization will imply an inflaton excursus of Planckian
values. Therefore, A future measurement of the $B$-mode polarization of the CMB will allow a 
determination of the value of the energy scale of inflation. 
This explains the utility of CMB polarization measurements 
as probes of the physics of inflation. A detection of primordial $B$-mode 
polarization would also demonstrate that inflation occurred at a very 
high energy scale, and that the inflaton traversed a super-Planckian 
distance in field space. 

%%%%%%%%%%%%%%%%%%%%%%%%%%%%%%%%%%%%%%%%%%%%%%%%%%%%%%%%%%%%%%%%
\begin{figure}
\centering\includegraphics[width=.5\linewidth]{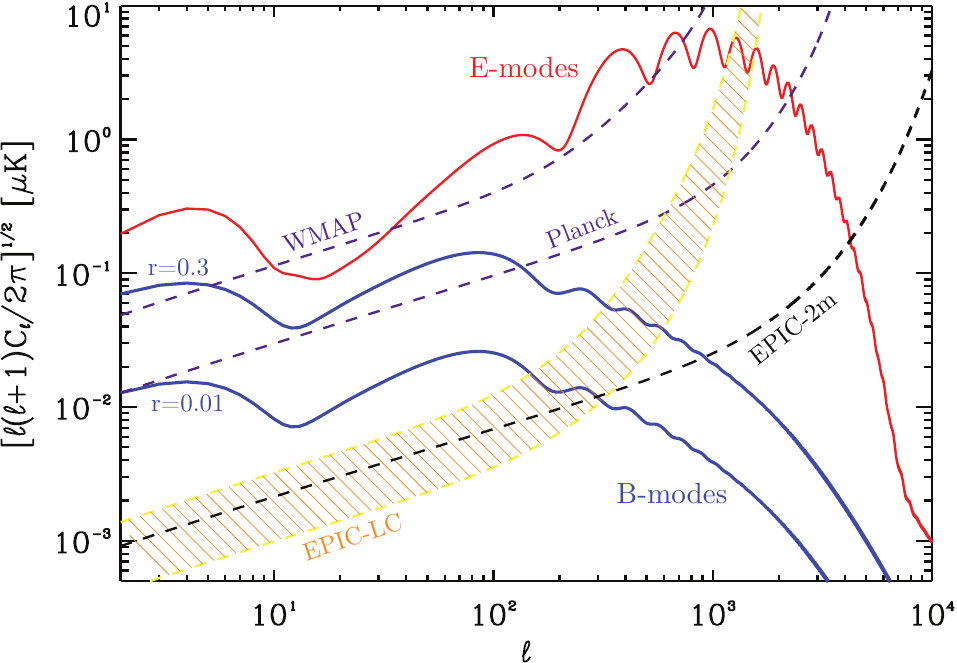}
\caption{E- and B-mode power spectra for a tensor-to-scalar ratio saturating the current bounds, $r=0.3$
and for $r=0.01$. Shown are the experimental sensitivities of WMAP, Planck and two different realizations
of CMBPol (EPIC-LC and EPIC-2m)}
\end{figure}
%%%%%%%%%%%%%%%%%%%%%%%%%%%%%%%%%%%%%%%%%%%%%%%%%%%%%%%%%%%%%%%%

\subsection{Dark matter}
Another topic we marginally touched in the lectures was dark matter, while we will not discuss the dark energy puzzle we did not have time to cover in detail.
This section is taken verbatim from Freese \cite{freese}. We thank K. Freese for granting permission.
The evidence that 95\% of the mass of galaxies and clusters is made of
some unknown component of Dark Matter (DM) comes from (i) rotation
curves (out to tens of kpc), (ii) gravitational lensing (out to 200
kpc), and (iii) hot gas in clusters. They lead us to believe that DM makes up about
30\% of the entire energy of the universe.

In the 1970s, Ford and Rubin  
discovered that rotation curves of
galaxies are flat.  The velocities of objects (stars or gas) orbiting
the centres of galaxies, rather than decreasing as a function of the
distance from the galactic centres as had been expected, remain
constant out to very large radii.  Similar observations of flat
rotation curves have now been found for all galaxies studied,
including our Milky Way.  The simplest explanation is that galaxies
contain far more mass than can be explained by the bright stellar
objects residing in galactic disks.  This mass provides the force to
speed up the orbits.  To explain the data, galaxies must have enormous
dark haloes made of unknown matter. Indeed, more than 95\% of
the mass of galaxies consists of dark matter.  The baryonic
matter which accounts for the gas and disk cannot alone explain the
galactic rotation curve. However, adding a DM halo allows a
good fit to data.

The limitations of rotation curves are that one can only look out as
far as there is light or neutral hydrogen (21 cm), namely to distances
of tens of kpc.  Thus one can see the beginnings of DM haloes, but
cannot trace where most of the DM is. The lensing experiments
discussed in the next section go beyond these limitations.

Einstein's theory of General Relativity predicts that mass bends, or
lenses, light.  This effect can be used to gravitationally ascertain
the existence of mass even when it emits no light.  Lensing
measurements confirm the existence of enormous quantities of DM
 both in galaxies and in clusters of galaxies.
Observations are made of distant bright objects such as galaxies or
quasars.  As the result of intervening matter, the light from these
distant objects is bent towards the regions of large mass.  Hence
there may be multiple images of the distant objects, or, if these
images cannot be individually resolved, the background object may
appear brighter.  Some of these images may be distorted or sheared.
The Sloan Digital Sky Survey used weak lensing (statistical studies of
lensed galaxies) to conclude that galaxies, including the Milky Way,
are even larger and more massive than previously thought, and require
even more DM out to great distances.  Again, the
predominance of DM in galaxies is observed.
The key success of the lensing of DM to date is the evidence that DM
is seen out to much larger distances than could be probed by rotation
curves: the DM is seen in galaxies out to 200 kpc from the centres
of galaxies, in agreement with
N-body simulations.  On even larger Mpc scales, there is
evidence for DM in filaments (the cosmic web).
Another piece of gravitational evidence for DM is the hot gas
in clusters.   The X-ray data indicates the presence of hot
gas.  The existence of this gas in the cluster can only be explained
by a large DM component that provides the potential well to
hold on to the gas.
In summary, the evidence is overwhelming for the existence of an
unknown component of DM that comprises 95\% of the mass in galaxies
and clusters.

There is another basic reason why DM is necessary: to form structures as we observe them. 
Let us assume that the matter content of the universe is dominated by
a pressureless and self-gravitating fluid. This approximation holds if
we are dealing with the evolution of the perturbations in the 
DM component or in case we are dealing with structures whose
size is much larger than the typical Jeans scale length of
baryons. Let us also define $\bx$ to be the co-moving coordinate and
${\bf r} = a(t){\bf x}$ the proper coordinate, $a(t)$ being the cosmic
expansion factor. Furthermore, if ${\bf v} =\dot{\bf r}$ is the physical
velocity, then ${\bf v} =\dot a{\bf x}+{\bf u}$, where the first term describes
the Hubble flow, while the second term, ${\bf u}=a(t)\dot{\bf x}$, gives the
peculiar velocity of a fluid element which moves in an expanding
background.

In this case the equations that regulate the Newtonian description of
the evolution of density perturbations are the continuity equation:
\be
{\partial \delta\over \partial t}+\nabla \cdot [(1+\delta){\bf u}]=0\,,
\label{eq:cont}
\ee
which gives the mass conservation, the Euler equation
\be
{\partial {\bf u}\over \partial t} + 2H(t) {\bf u}+ ({\bf u}\cdot\nabla){\bf u} =
-{\nabla \phi\over a^2}\,,
\label{eq:eul}
\ee
which gives the relation between the acceleration of the fluid element
and the gravitational force, and the Poisson equation
\be
\nabla^2\phi = 4\pi G\bar\rho a^2\delta
\label{eq:pois}
\ee
which specifies the Newtonian nature of the gravitational force. In
the above equations, $\nabla$ is the gradient computed with
respect to the co-moving coordinate ${\bf x}$, $\phi({\bf x})$ describes the
fluctuations of the gravitational potential, and $H(t)=\dot a/a$ is the
Hubble parameter at the time $t$. Its time-dependence is given by
$H(t)=E(t)H_0$, where 
\be
E(z)=[(1+z)^3\Omega_m+(1+z)^2(1-\Omega_m-\Omega_{DE})+(1+z)^{3(1+w)}
\Omega_{DE}]^{1/2}. 
\label{eq:ez}
\ee

In the case of small perturbations, these equations can be linearized
by neglecting all the terms which are of second order in the fields
$\delta$ and ${\bf u}$.
 In this case, using the Euler equation to
eliminate the term $\partial \bu/\partial t$, and using the Poisson
equation to eliminate $\nabla^2\phi$, one ends up with
\be
{\partial^2 \delta\over \partial t^2}+2H(t){\partial
  \delta\over \partial t}-4\pi G\bar\rho\delta=0\,.
\label{eq:linear}
\ee
This equation describes the Jeans instability of a pressureless fluid,
with the additional ``Hubble drag'' term $2H(t) {\partial \delta /
\partial t}$, which describes the counter-action of the expanding
background on the perturbation growth. Its effect is to prevent the
exponential growth of the gravitational instability taking place in a
non-expanding background.  The solution
of the above equation can be cast in the form:
\be
\delta({\bf x},t) = \delta_+({\bf x},t_i)D_+(t)+\delta_-({\bf x},t_i)D_-(t)\,,
\label{eq.linsol}
\ee
where $D_+$ and $D_-$ describe the growing and decaying modes of the
density perturbation, respectively. In the case of an
Einstein--de-Sitter (EdS) universe ($\Omega_m=1$, $\Omega_{DE}=0$), it
is $H(t)=2/(3t)$, so that $D_+(t)=(t/t_i)^{2/3}$ and $D_-(t)=
(t/t_i)^{-1}$. The fact that $D_+(t)\propto a(t)$ for an EdS universe
should not be surprising. Indeed, the dynamical time-scale for the
collapse of a perturbation of uniform density $\rho$ is $t_{\rm
dyn}\propto (G\rho)^{-1//2}$, while the expansion time-scale for the
EdS model is $t_{\rm exp}\propto (G\bar\rho)^{-1//2}$, where
$\bar\rho$ is the mean cosmic density. Since for a linear (small)
perturbation it is $\rho \simeq \bar\rho$, then $t_{\rm dyn}\sim
t_{\rm exp}$, thus showing that the cosmic expansion and the
perturbation evolution take place at the same pace. This argument also
leads to understanding the behaviour for a $\Omega_m<1$ model. In this
case, the expansion time scale becomes shorter than the above one at
the redshift at which the universe recognizes that $\Omega_m<1$. This
happens at $1+z\simeq \Omega_m^{-1/3}$ or at $1+z\simeq \Omega_m^{-1}$
in the presence or absence of a cosmological constant term,
respectively. Therefore, after this redshift, cosmic expansion takes
place at a quicker pace than gravitational instability, with the
result that the perturbation growth is frozen.

The exact expression for the growing model of perturbations is given
by 
\be
D_+(z)\,=\,{5\over 2}\,\Omega_m E(z)\,\int_z^\infty {1+z'\over
E(z')^3}\, dz'.
\label{eq:grw}
\ee 
The EdS has the faster evolution, while the slowing down
of the perturbation growth is more apparent for the open low-density
model, the presence of a cosmological constant providing an intermediate
degree of evolution. 
The key point is, however, that a pressureless fluid such as DM is needed for the perturbations to grow to
give rise to collapsed objects. Baryon perturbations, being coupled to photons till the last-scattering  epoch, 
feel a non-vanishing pressure and therefore they may not grow. After the last-scattering stage, the baryons 
fall into the gravitational 
potential generated by DM and the baryonic perturbations may promptly catch up with those of DM.

\subsubsection{Dark matter candidates}
 
 There is a plethora of dark matter candidates. 
MACHOs, or Massive Compact Halo Objects, are made of
ordinary matter in the form of faint stars or stellar remnants;
they could also be primordial black holes or mirror matter.
However, there are not enough of them to completely resolve the
question.  Of the non-baryonic candidates, the most popular are the
WIMPS (Weakly Interacting Massive Particles) and the axions, as these
particles have been proposed for other reasons in particle physics.
Ordinary massive neutrinos are too light to be cosmologically
significant, though sterile neutrinos remain a possibility.  Other
candidates include primordial black holes, non-thermal WIMPzillas, and
Kaluza--Klein particles which arise in higher dimensional theories.

About axions, the good news is that cosmologists do not need to ``invent'' new
particles.  Two candidates already exist in particle physics for other
reasons: axions and WIMPs.  Axions with masses in the range
$10^{-(3-6)}$ eV arise in the Peccei--Quinn solution to the strong-CP
problem in the theory of strong interactions.  

WIMPs are also natural dark matter candidates from particle physics.
These particles, if present in thermal abundances in the early
universe, annihilate with one another so that a predictable number of
them remain today.  The relic density of these particles comes out to
be the right value:
\begin{equation}
\Omega_{\rm DM} h^2 = (3 \times 10^{-26} {\rm cm}^3/{\rm s})
/ \langle \sigma v \rangle_{\rm A}
\end{equation}
where the annihilation cross-section $\langle \sigma v \rangle_{\rm A} $
of weak interaction strength automatically gives the right answer. The reason why
the final abundance is inversely proportional to the annihilation cross-section is rather clear: the larger the
annihilation cross-section, the more WIMPs annihilate and the fewer of them are left behind. Furthermore, annihilation
is not eternal: owing to the expansion of the universe, annihilation stops when its rate becomes smaller than the
expansion rate of the universe. When this happens,  the abundance is said to freeze-out.

\begin{figure}
\centering\includegraphics[width=.5\linewidth]{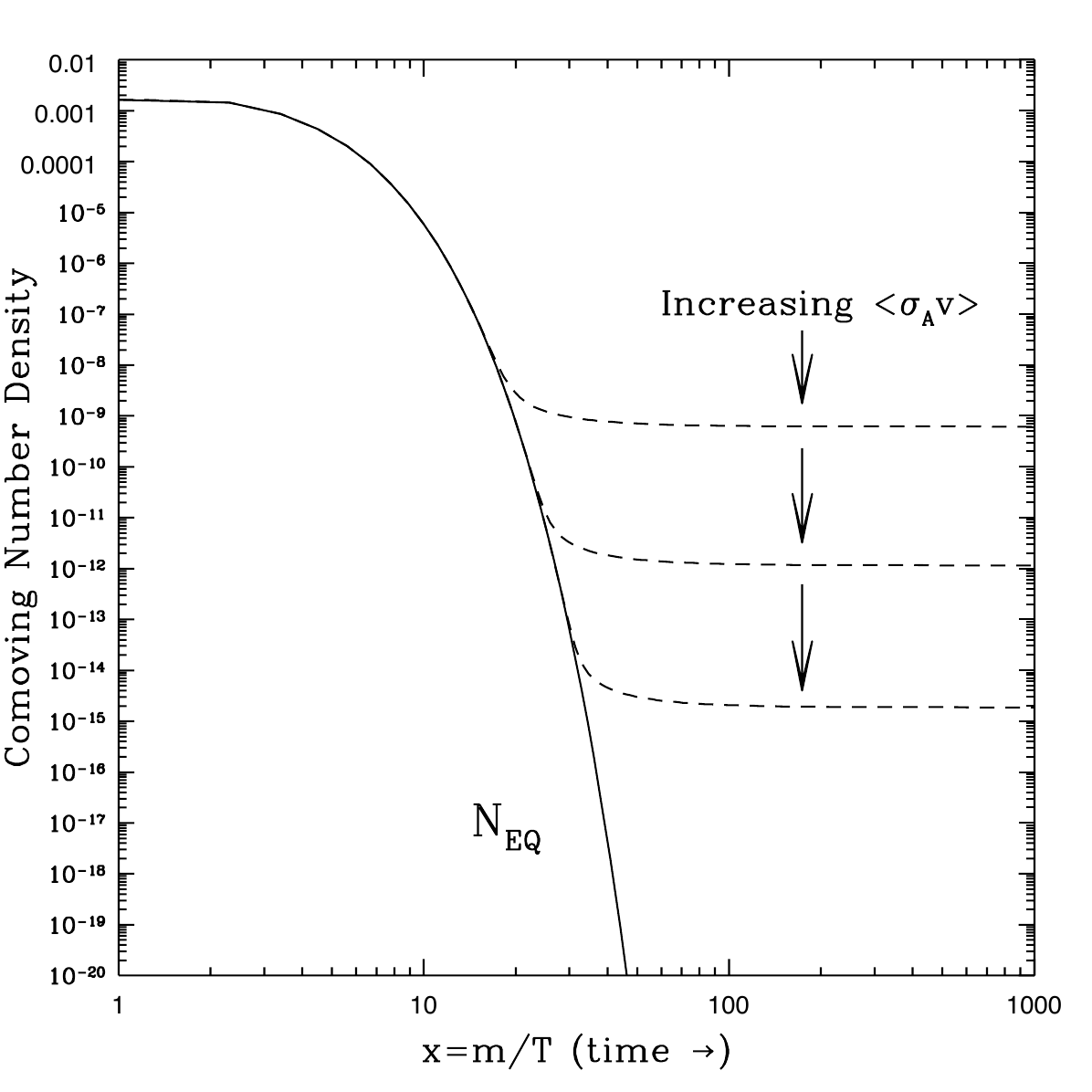}
\caption{The abundance of WIMPs of a given mass $m$ as a function of temperature and for various
annihilation cross-sections}
\end{figure}

This coincidence is known as `the WIMP miracle' and is the reason why
WIMPs are taken so seriously as DM candidates.  The best WIMP
candidate is motivated by Supersymmetry (SUSY): the lightest
neutralino in the Minimal Supersymmetric Standard Model.
Supersymmetry in particle theory is designed to keep particle masses
at the right value.  As a consequence, each particle we know has a
partner: the photino is the partner of the photon, the squark is the
quark's partner, and the selectron is the partner of the electron.
The lightest superysmmetric partner is a good dark matter candidate.

There are several ways to search for dark WIMPs. SUSY particles may be
discovered at the LHC as missing energy in an event.  In that case one
knows that the particles live long enough to escape the detector, but
it will still be unclear whether they
are long-lived enough to be the dark matter.  Thus
complementary astrophysical experiments are needed. In direct
detection experiments, the WIMP scatters off  a nucleus in the
detector, and a number of experimental signatures of the interaction
can be detected.  In indirect detection experiments,
neutrinos  that arise as
annihilation products of captured WIMPs exit from the Sun and can be detected on Earth.  Another way to detect WIMPs
is to look for anomalous cosmic rays from the Galactic Halo: WIMPs in
the Halo can annihilate with one another to give rise to antiprotons,
positrons, or neutrinos.  In addition,
neutrinos, gamma rays, and radio waves may be detected as WIMP
annihilation products from the Galactic Centre. For lack of time these issues 
were not discussed extensively  in the lectures. The interested reader may find more
about these issues in Ref. \cite{bertone}.

\section{Conclusions}
The period when we say that cosmology is entering a golden age has already passed: cosmology {\it is} in the 
middle of  its golden age.
Present observational data pose various puzzles  whose solutions might either be around the corner   
or decades far in the future. It will require some young and creative researcher sitting in this 
room to solve them. This is why the cosmological puzzles are dark, but the future is brighter.

\section*{Acknowledgements}
It is a great pleasure to thank all the
organizers, N. Ellis, E. Lillistol, D. Metral, and especially M. Losada and
E. Nardi, for having created such a stimulating atmosphere. All students
are also acknowledged for their never-ending enthusiasm. This version contains hopefully the right citations to other lectures notes by  K. Freese, E.W. Kolb, A. Linde, M. Turner  and M. Zaldarriaga which the author  used verbatim. We apologize with them for this admittedly late note.

\end{document}